\documentclass{article}
\usepackage{arxiv}
\usepackage{booktabs}
\usepackage[pdftex]{graphicx}
\usepackage{color, colortbl}
\usepackage{subcaption}
\usepackage{fixltx2e}
\usepackage{url}
\usepackage{amsmath}
\usepackage{mathtools}
\usepackage[disable]{todonotes}
\usepackage{cleveref}
\usepackage{verbatim}
\usepackage{amsmath}
\usepackage{graphicx}
\usepackage{tabu}
\usepackage{multirow}
\usepackage{lipsum}
\usepackage{flushend}
\usepackage{mdframed}
\usepackage{textcomp}
\usepackage{float}
\usepackage{xspace}
\usepackage{fancybox}

\newcommand{\ie}{i.e.,\xspace}
\newcommand{\eg}{e.g.,\xspace}

\newcommand{\etal}{et al.\xspace}

\newcommand{\model}[1]{{\fontfamily{cmss}\selectfont #1}}
\newcommand{\CCOTF}{\model{RNN\textsubscript{\textsc{CCotf}}}\xspace}
\newcommand{\CC}{\model{RNN\textsubscript{\textsc{CC}}}\xspace}
\newcommand{\SOAPI}{\model{RNN\textsubscript{\textsc{SOapi}}}\xspace}
\newcommand{\SO}{\model{RNN\textsubscript{\textsc{SO}}}\xspace}
\newcommand{\MFONE}{\model{MF1}\xspace}
\newcommand{\MFTWO}{\model{MF2}\xspace}

\newcommand{\lstm}{\textsc{LSTM}\xspace}
\newcommand{\cado}{\textsc{CaDO}\xspace}
\newcommand{\python}{\textsc{Python}\xspace}
\newcommand{\scumble}{\textsc{SCUMBLE}\xspace}
\newcommand{\rnn}{\textsc{RNN}\xspace}
\begin{document}
\title{On Using Machine Learning to Identify Knowledge in API Reference Documentation}
\author{
 Davide Fucci \\
  University of Hamburg\\
  Hamburg\\
  Germany \\
  \texttt{fucci@informatik.uni-hamburg.de} \\
  %% examples of more authors
   \And
 Alireza Mollaalizadehbahnemiri \\
  University of Hamburg\\
  Hamburg\\
  Germany \\
  \texttt{alirezam.alizadeh@gmail.com} \\
   \And
   Walid Maalej\\
   University of Hamburg\\
   Hamburg\\
   Germany\\
   \texttt{maalej@informatik.uni-hamburg.de}
  %% Coauthor \\
  %% Affiliation \\
  %% Address \\
  %% \texttt{email} \\
  %% \And
  %% Coauthor \\
  %% Affiliation \\
  %% Address \\
  %% \texttt{email} \\
  %% \And
  %% Coauthor \\
  %% Affiliation \\
  %% Address \\
  %% \texttt{email} \\
}

% \author{\IEEEauthorblockN{Anonymous}
% \IEEEauthorblockA{School of Electrical and\\Computer Engineering\\
% Georgia Institute of Technology\\
% Atlanta, Georgia 30332--0250\\
% Email: http://www.michaelshell.org/contact.html}
% }

% \and
% \IEEEauthorblockN{Anonymous}
% \IEEEauthorblockA{Twentieth Century Fox\\
% Email: homer@thesimpsons.com}
% \and
% \IEEEauthorblockN{Anonymous}
% \IEEEauthorblockA{Starfleet Academy\\
% San Francisco, California 96678--2391\\
% Telephone: (800) 555--1212\\

\maketitle

\begin{abstract}
Using API reference documentation like JavaDoc is an integral part of software development.
Previous research introduced a grounded taxonomy that organizes API documentation knowledge in 12 types, including knowledge about the \textit{Functionality}, \textit{Structure}, and \textit{Quality} of an API. 
We study how well modern text classification approaches can automatically identify documentation containing specific knowledge types.
We compared conventional machine learning (\textit{k}-NN and SVM) and deep learning approaches trained on manually-annotated Java and .NET API documentation ($n$ = 5,574). 
When classifying the knowledge types individually (i.e., multiple binary classifiers) the best AUPRC was up to 87\%. 
The deep learning and SVM classifiers seem complementary.
For four knowledge types (\textit{Concept}, \textit{Control}, \textit{Pattern}, and \textit{Non-Information}), SVM clearly outperforms deep learning which, on the other hand, is more accurate for identifying the remaining types.
When considering multiple knowledge types at once (i.e.,  multi-label classification) deep learning outperforms na\"ive baselines and traditional machine learning achieving a MacroAUC up to 79\%. 
We also compared classifiers using embeddings pre-trained on generic text corpora and StackOverflow but did not observe significant improvements.
Finally, to assess the generalizability of the classifiers, we re-tested them on a different, unseen Python documentation dataset. 
Classifiers for \textit{Functionality}, \textit{Concept}, \textit{Purpose}, \textit{Pattern}, and \textit{Directive} seem to generalize from Java and .NET to Python documentation.  
The accuracy related to the remaining types seems  API-specific.
We discuss our results and how they inform the development of tools for supporting developers sharing and accessing API knowledge.
Published article: \url{https://doi.org/10.1145/3338906.3338943}
\end{abstract}

\section{Introduction}
Software developers reuse software libraries and frameworks through Application Programming Interfaces (APIs).  
They often rely on reference documentation to identify which API elements are relevant for the task at hand, how the API can be instantiated, configured, and combined~\cite{RD10}. 
Compared to other knowledge sources, such as tutorials and Q\&A portals, reference documentation like JavaDoc and PyDoc are considered the official API technical documentation.
They provide detailed and fundamental information about API elements, components, operations, and structures~\cite{DH09,MR13}.

As API documentation can be thousands of pages long~\cite{PRD15,RC14}, accessing relevant knowledge can be tedious and time-consuming~\cite{RD10}.
Moreover, the information necessary to accomplish a task can be scattered across the documentation pages of multiple  elements, such as classes, methods, and properties. 
Thus, developers try to use other sources to fulfill their information needs.
For example, although the Java Development Kit (JDK) API documentation contains more than 7,000 pages, as of early 2019, there are more than 3 million StackOverflow posts tagged as \texttt{java}.

\begin{figure}
    \centering    
    \frame{\includegraphics[width=.48\textwidth]{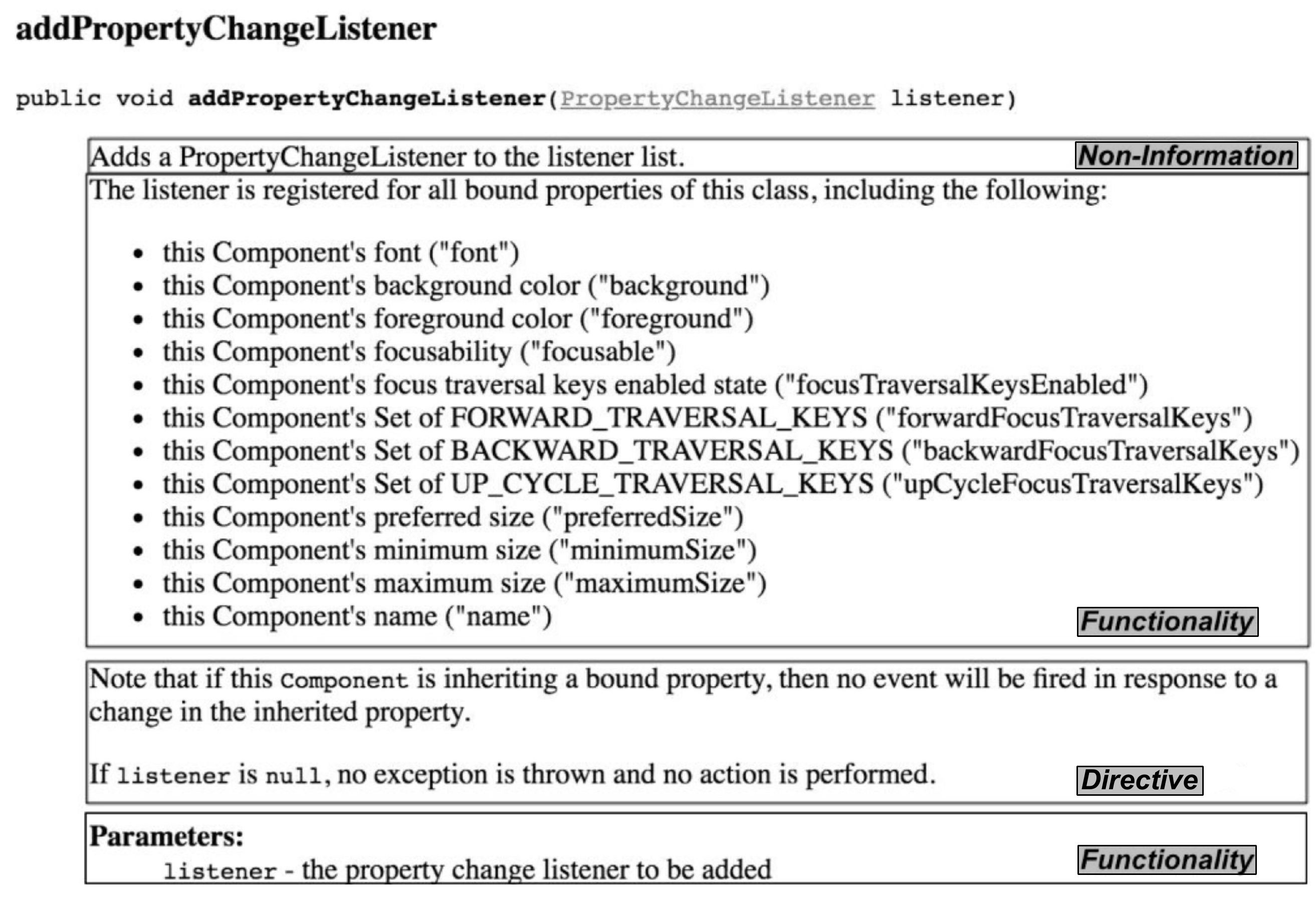}}
    \caption{A reference documentation page in the JDK API annotated with the knowledge types it contains.}
    \label{fig:example1}
\end{figure}

Over the last decade, software engineering researchers studied what information developers need when consulting API documentation~\cite{MR13,SGB08, SM08}.
One line of research focuses on automatically matching information needs with the types of knowledge available in the documentation.
Maalej and Robillard~\cite{MR13} took a first step in this direction by developing an empirically-validated taxonomy of 12 knowledge types found within API reference documentation.
% The taxonomy is composed of 12 knowledge types, including a \textit{non-information} type---\ie boilerplate, uninformative text.
A single documentation page can include several knowledge types (\Cref{fig:example1}).
\textit{Functionality} and  \textit{Directive} are particular types of knowledge needed to accomplish a development task, whereas the \textit{Non-information} type contains only uninformative boilerplate text~\cite{MR13}. 
Maalej and Robillard argue that such knowledge categorization allows for a) understanding and improving the documentation quality and b) satisfying developers' information needs.~\cite{MR13} 

The research community has shown interest in studying specific knowledge types contained in API reference documentation.  
For example, Montperrus et al.~\cite{MET11} and Seid et al.~\cite{SSD15} studied \textit{Directive} to prevent the violation of API usage constraints. 
Robillard and Chhetri~\cite{RC14} filtered \textit{Non-information} when recommending APIs to developers.
However, these automated approaches are based either on linguistic features engineering~\cite{RC14} or on syntactic patterns~\cite{MET11}.   

This work investigates how well modern text classification approaches can automatically identify the knowledge types suggested by Maalej and Robillard in API documentation.
Based on a dataset of 5,574 labelled Java and .NET documentation, we trained, tested, and compared conventional machine learning approaches---i.e., \textit{k}-Nearest Neighbors (\textit{k}-NN) and Support Vector Machines (SVM)---as well as deep learning approaches---i.e., recurrent neural network (RNN) with a Long Short-Term Memory (LSTM) layer. 
The RNN learns features from a semantic representation of general purpose text (i.e.,~embeddings).
Hence, we studied how our results are impacted by training the network using software development-specific corpora from StackOverflow as opposed to a general purpose one. 
Finally, we studied the generalizability of the classifiers to an unseen dataset obtained from the Python standard library.
% Our results reveal that the generalization of the classifiers are limited to a few knowledge type.

This paper makes three contributions.
First, we present a detailed classification  \textbf{benchmark} for API documentation. 
The settings include different machine learning approaches and configurations, different word embeddings for the RNN, different datasets for different APIs, as well as various evaluation metrics. 
Researchers and tool vendors can use the benchmark, for example, to select and optimize a specific classifier for a specific cofiguration of API and knowledge types. 
Second, as we share the code and data of this study,\footnote{\url{https://zenodo.org/badge/latestdoi/194706952
}} several \textbf{top-performing classifiers}  (e.g., AUPRC $\ge$ 80\%) already have practical relevance.
Third, our findings and discussion of related work provide insights to researchers, tool vendors, and practitioners on \textbf{how machine learning can help} better organize, access, and share knowledge about API.  

% The goal of this study is to investigate to what extent the knowledge types defined in Maleej and Robillard~\cite{MR13} can be learned from the text semantic features of API reference documentation of the \@.NET and JDK framework and, afterwards, automatically identified.
% Our purpose it to understand the impact of using a textual representation trained on a technical corpus rather than on a general-purpose one. 
% Moreover, we are interested in assessing the applicability of our model to the identify the knowledge types within API reference documentation of other projects.
% In particular, with this study, we answer the following research questions (RQ): 
% \begin{itemize}
%     \item \rqs{1}{Can RNN be used to automatically identify knowledge types within API reference documentation?}

% \item \rqs{2}{To what extent the corpora used to train the RNN input layer impacts the identification of knowledge types in API reference documentation?}

%     \item \rqs{3}{Can our approach identify knowledge types within API reference documentation characterized by a different documentation style?}
% \end{itemize}

% \noindent\textbf{Paper organization.} 
%\Cref{sec:background} reports background information concerning the problem tackled in this study as well as the RNN used.
The rest of the paper is organized as follows. 
\Cref{sec:methodology} describes our research settings, \Cref{sec:configs} presents the configurations of the  classifiers, and \Cref{sec:results} reports their performance We discuss related work in \Cref{sec:related} and the implication of our results in \Cref{sec:discussion}. 
Finally, \Cref{sec:conclusion} concludes the paper.
\section{Research Settings}
\label{sec:methodology}
% This section reports the details of our approach.
This section introduces the research questions, method, and data.    

% Moreover, as each API reference document can contain several non-mutually exclusive types of knowledge, we run the same comparison for more challenging task of multi-label identification. 
% Our approach is based on an RNN with three hidden layers, one \lstm unit and two fully-connected layers.
%This network can learn long-term dependencies as in the case of a large text.
% One of the earliest automated approaches to classify the knowledge types in JAVA APIs reference documentation was proposed by Metkar and Kamalapur (2014). They introduced an automated system based on Natural Language Processing (NLP) to categorize the knowledge using pattern-based Information Retrieval (IR) techniques. Robillard and Chhetri (2015) proposed a recommander system which extracts the most important pieces of information from the APIs documentation based on IR techniques and supports the developers in understanding how to use an API by recommending the knowledge items related to that API during the development. Eventually, Kumar and Devanbu (2016) introduced a model based on Support Vector Machines (SVM) to perform the knowledge categorization in Python reference documents.

\subsection{Research Questions and Method}
%In this paper, we are interested in answering the following research questions:
Maalej and Robillard~\cite{MR13} proposed an empirically-validated taxonomy of 12 knowledge types based on grounded theory and systematic content analysis (17 experienced coders, 279 person-hours effort).  
\Cref{tbl:taxonomy} reports the identified knowledge types which represent the basis for this work.    
Our primary goal is to study how well simple machine learning for text classification, without additional feature engineering or advanced natural language processing (NLP) techniques, can identify these knowledge types.
That is, our classifiers label a document with one or more knowledge types. 

    \begin{table}[!ht]
        \centering
        \caption{Twelve knowledge types included in reference documentation (adapted from Maalej and Robillard~\cite{MR13}). }
        \begin{tabu} to \linewidth {lX}
        \toprule
        \multicolumn{1}{c}{Knowledge type} & \multicolumn{1}{c}{Brief description}                                                    \\ \midrule
        Functionality                                     & Describes the capabilities of the API, and what happens when it is used.           \\
        \rowcolor[HTML]{EFEFEF} 
        Concept                                           & Explains terms used to describe the API behavior or the API implementation.       \\
        Directive                                         & Describe what the user is allowed (not allowed) to do with the API.                \\
        \rowcolor[HTML]{EFEFEF} 
        Purpose                                           & Explains the rationale for providing the API or for a design decision.             \\
        Quality                                           & Describes non-functional attributes of the API, including its implementation.      \\
        \rowcolor[HTML]{EFEFEF} 
        Control                                           & Describes how the API manages the control-flow and sequence of calls.              \\
        Structure                                         & Describes the internal organization of API elements including their relationships. \\
        \rowcolor[HTML]{EFEFEF} 
        Pattern                                           & Explains how to get specific results using the API.                              \\
        Example                                           & Provides examples about the API usage.                                 \\
        \rowcolor[HTML]{EFEFEF} 
        Environment                                       & Describes the API usage environment.                                \\
        Reference                                         & Pointers to external documents.                                                    \\
        \rowcolor[HTML]{EFEFEF} 
        Non-information                                   & Uninformative, boilerplate text.                                                   \\ \bottomrule
        \end{tabu}
        \label{tbl:taxonomy}
 \end{table}
 
There are two main text classification approaches which we study in this paper. 
Traditional approaches usually learn the classes from the occurrences of certain keywords or phrases in the training set. 
More computational intensive approaches, often referred to as deep learning, use the semantics of the keywords---i.e., the context of the keyword occurrences~\cite{DY14}.

For traditional approaches, we study two algorithms frequently used for text classification, \textit{k}-NN and SVM. 
For deep learning, we used RNN with an LSTM layer, which is particularly effective for text categorization problems~\cite{HS17}. 
This architecture is recommended over, for example, Convolutional Neural Network (CNN).
While the latter is more suited for image recognition~\cite{LBH15}, RNN with LSTM handles more efficiently the dependencies between features~\cite{HS17}.
We also compare these classifiers to na\"ive baselines.

The task tackled in this study is to assign knowledge types to an API document. 
As the document can contain more than one knowledge type, this task is modelled as a multiple binary classification problem consisting of independently train one binary classifier for each knowledge type.  
Another approach is to train a multi-label classifier---i.e., a classifier that outputs a set of knowledge types rather than a single one. 
We analyze and report the results for both approaches when answering the following research question.
\begin{mdframed}[backgroundcolor=lightgray!20,skipabove=5pt]
    \item \textbf{RQ1}.~How well can text-based classifiers identify knowledge types in API reference documentation? In particular, can deep learning improve over traditional approaches?
\end{mdframed}
\vspace{5pt}
For text classification tasks, the input layer of an RNN usually consists of embeddings trained on large unlabeled textual corpora necessary to capture rich semantic features~\cite{MSC13}. 
Pre-trained embeddings are available and can be easily ``plugged'' in the network without further effort. 
However, while these embeddings save computational time and well represent common language tasks, they can miss software engineering or API-specific semantics. 
This motivates our second research question.
\begin{mdframed}[backgroundcolor=lightgray!20,skipabove=5pt,skipbelow=10pt]
    \item \textbf{RQ2}. Do software development-specific text embeddings improve classification results compared to general purpose ones?
\end{mdframed}
\vspace{5pt}
% Moreover, we create two additional embeddings using text scraped from StackOverflow and use them as input layer for the LSTM.  
Finally, a common question for machine learning evaluation is whether a model trained on a certain dataset generalizes to other data. 
The original dataset includes documentation of the standard Java and .NET libraries~\cite{MR13}. 
Since we aim to assess the generalizability of our approach to API reference documentation written in a different style, we manually annotated a new datase sampled from the Python standard library documentation.
We used this dataset as an additional test set to report our classifiers performance.
\begin{mdframed}[backgroundcolor=lightgray!20,skipabove=5pt,skipbelow=0pt]
     \item\textbf{RQ3}. Can documentation classification based on knowledge types be generalized across API?
\end{mdframed}
\vspace{5pt}
% \usepackage{booktabs}
% \usepackage{booktabs}

% \subsection{Research Method}
We assess models based on 10-fold cross-validation using 10\% of the dataset as test set.
When comparing individual knowledge types classifiers, we report Area Under Precision-Recall Curve (AUPRC).
Precision-Recall curves are a common metric to evaluate binary classification and are obtained by plotting precision and recall values at different probabilities thresholds~\cite{BEP13}.
In particular, they are used to evaluate machine learning model trained on imbalanced data sets~\cite{BEP13}.
Therefore, AUPRC is a summary measure of performance irrespective of a particular threshold.

When comparing classifiers for multiple knowledge types, we report performance according to two types of metrics, item-based and label-based.
The item-based metrics are a)
% \begin{itemize}
\textit{Hamming Loss}, namely the ratio of wrongly classified labels to the total number of labels (its best value is zero) and b) 
\textit{Subset Accuracy}, namely the percentage of exact matches between the predicted and the actual labelset.

% \end{itemize} 
The label-based metrics are precision, recall, F1-measure (Formula \ref{eq:f1}), and Area Under Receiving Operator Curve (AUC).
 \begin{equation}
\footnotesize
 F1 = \frac{2 \times True Positives}{2 \times True Positives + False Positives + False Negatives}
 \label{eq:f1}
\end{equation}
The Receiving Operator Curve is created by plotting recall against false positive rate (FPR, Equation~\ref{eq:specificity}), at different probability thresholds.
Accordingly, AUC does not depend on a particular threshold~\cite{HL05}.
To calculate the value of True Positive, False Positives, and False Negatives we used 0.5 as probability threshold~\cite{DY14}.
\begin{equation}
\footnotesize
FPR = 1 - \frac{True Negatives}{True Negatives + False Positives} 
 \label{eq:specificity}
\end{equation}

The label-based metrics are macro-averaged.
Macro-averaging applies the metric to the binary partition of each predicted label and then averages the results---i.e., labels have equal contribution in the final result. 
In contrast, micro-averaging first aggregates the individual metric components (i.e., true positives, false positives, true negatives, and false negatives) of each label and then averages them.
Therefore, micro-averaging is biased toward the majority classes and should be avoided when evaluating unbalanced datasets~\cite{SL09}. 

We compare the results of the classifiers to na\"ive baselines, 
\MFONE, \MFTWO, and RAND. 
The first two always assign the first (respectively one of the first two) most-frequent labels to each document, whereas the latter assigns a random label. 

\begin{table}[ht!]
    \centering
    \caption{Overview of the \textsc{CaDO} dataset.}
    \begin{tabular}{lrrrr}
    \toprule
          & \#documents  & Words max.  & Words mean   & Vocab. size \\ \midrule
    .NET  & 2,782             & 2,874 & 89   & 10,630       \\
    JDK   & 2,792             & 2,099 & 86   & 10,763       \\ \midrule
    Total & 5,574             & 2,874 & 87   & 17,758       \\ \bottomrule
    \end{tabular}
    \label{tbl:dataset}
    \end{table}

\subsection{Research Data}
We use the \textbf{\cado dataset} created by Maalej and Robillard~\cite{MR13} as the result of their content analysis of the JDK 6 and .NET 4.0 API reference documentation.
\cado contains 5,574 observations.
The columns include the name of the API element (e.g., a class, a method, or a property), its documentation text, and 12 binary values indicating the presence (or absence) of the corresponding knowledge type. 
~\Cref{tbl:dataset} summarizes the dataset textual properties.

The most frequent knowledge types are \textit{Functionality} and \textit{Non-information}, whereas \textit{Quality} and \textit{Environment} are the least frequent.
We did not merge some of the knowledge types as Maalej and Robillard reported no significant evidence of their co-occurence~\cite{MR13}.

The majority of the documents (90.5\%) contains one to five of the 12 knowledge types. 
We use the \scumble~\cite{CRD14} score ($\in$ [0, 1]) to report the level of unbalancedness.
For a given label, a high \scumble score represents a large difference between the frequencies of all the other co-occurring labels.
In general, datasets with high scores are problematic for classification tasks~\cite{HCR16}. 
However, for datasets characterized by low \scumble score, resampling can reduce unbalancedness~\cite{HCR16}.
\cado mean \scumble score is 0.11.

\begin{figure}[!b]
      \centering
      \includegraphics[width=\linewidth]{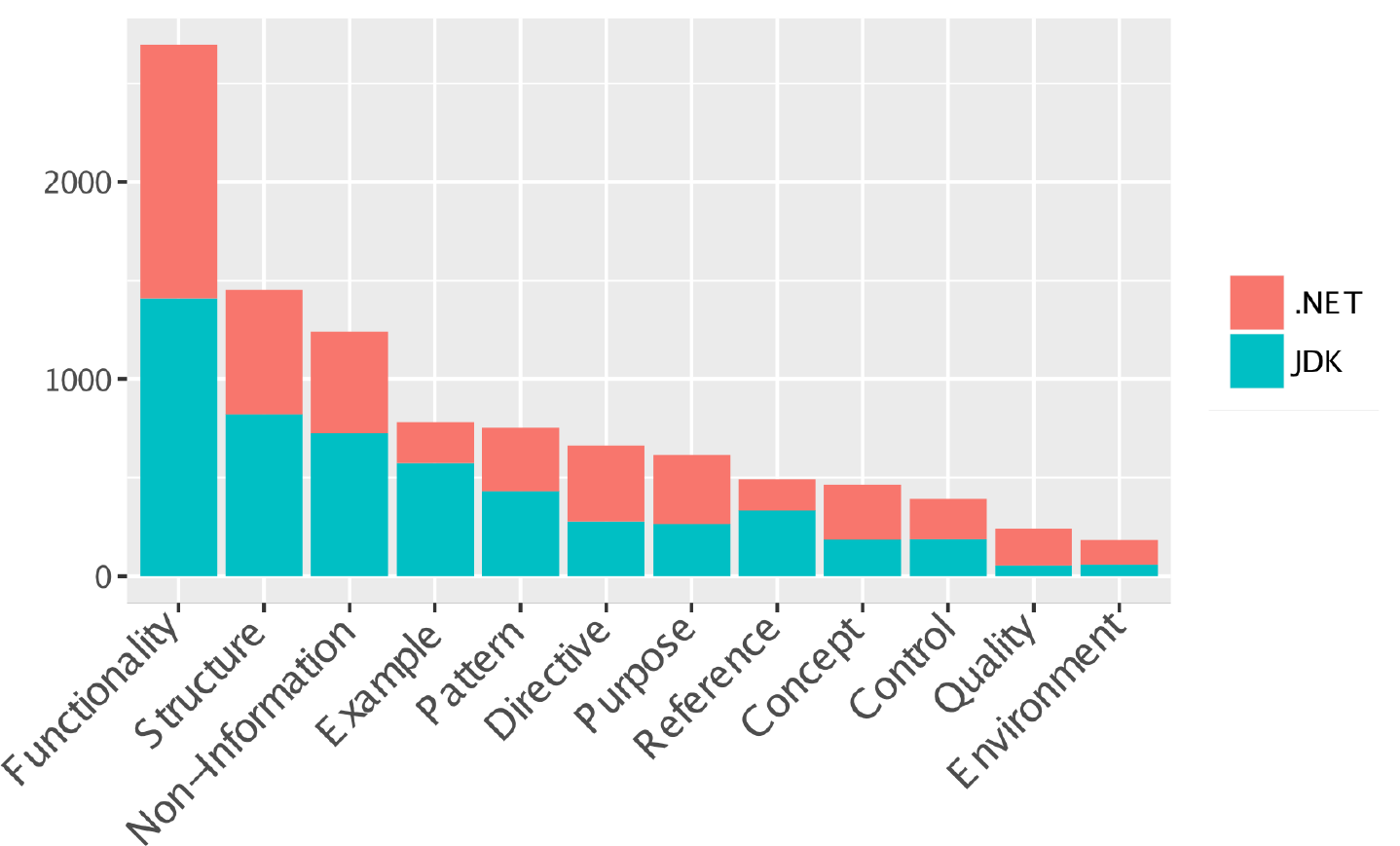}
      \caption{Knowledge types distribution in the \cado dataset after resampling.}
        \label{fig:datasets}
\end{figure}

We applied random under- and over-sampling to 90\% of the dataset (i.e., the training set).
We did not resample the test set (10\% of the dataset) to avoid sampling bias.
For resampling, we removed 30\% of the documents containing \textit{Functionality} and \textit{Non-Information} in their labelset and duplicated 50\% of the documents containing \textit{Environment} and \textit{Quality}. The thresholds were obtained empirically based on the \scumble score. 
After resampling, the training and test sets contain 3,876 and 430 observations respectively.
\Cref{fig:datasets} presents the label frequencies in the dataset we used to train the models after re-sampling.

% \begin{figure*}
%     \begin{subfigure}{.5\textwidth}
%       \centering
%       \includegraphics[width=\linewidth]{fig/original_cado}
%       \caption{Original (unbalanced) \cado dataset \cite{MR13}.}
%       \label{fig:sfig1}
%     \end{subfigure}%
%     \begin{subfigure}{.5\textwidth}
%       \centering
%       \includegraphics[width=\linewidth]{fig/balanced_cado}
%       \caption{\cado dataset after resampling to address unbalancedness.}
%       \label{fig:sfig2}
%     \end{subfigure}
%     \caption{Knowledge types distribution in the \cado dataset.}
%         \label{fig:datasets}
% \end{figure*}
We prepared a new \textbf{\python dataset} consisting of 100 API documentation pages (\ie modules, types, attributes, and methods) from the Python 2.7 standard library.\footnote{https://docs.python.org/2.7/library/}
We selected the Python standard library since its code is organized differently than Java or~.NET as it makes extensive use of modules in which functions, classes, and variables are defined.
The Python programming paradigm is more functional than Java and~.NET which instead follow an object-oriented paradigm. 
Python is dynamically typed, and its reference documentation tends to focus on functions, whereas types documentation is embeddable in the source code (e.g., through Docstrings).
Finally, its development and documentation are driven by an open source, non-profit community (the Python Software Foundation) whereas Java and~.NET are owned by corporations.

We followed the sampling strategy suggested by Maalej and Robillard~\cite{MR13}---i.e., stratified random sampling. 
We first created strata for each of the base modules and then randomly sampled API documentation from each stratum proportionally to their frequencies. 

\begin{figure}[t]
    \centering
    \includegraphics[width=0.48\textwidth]{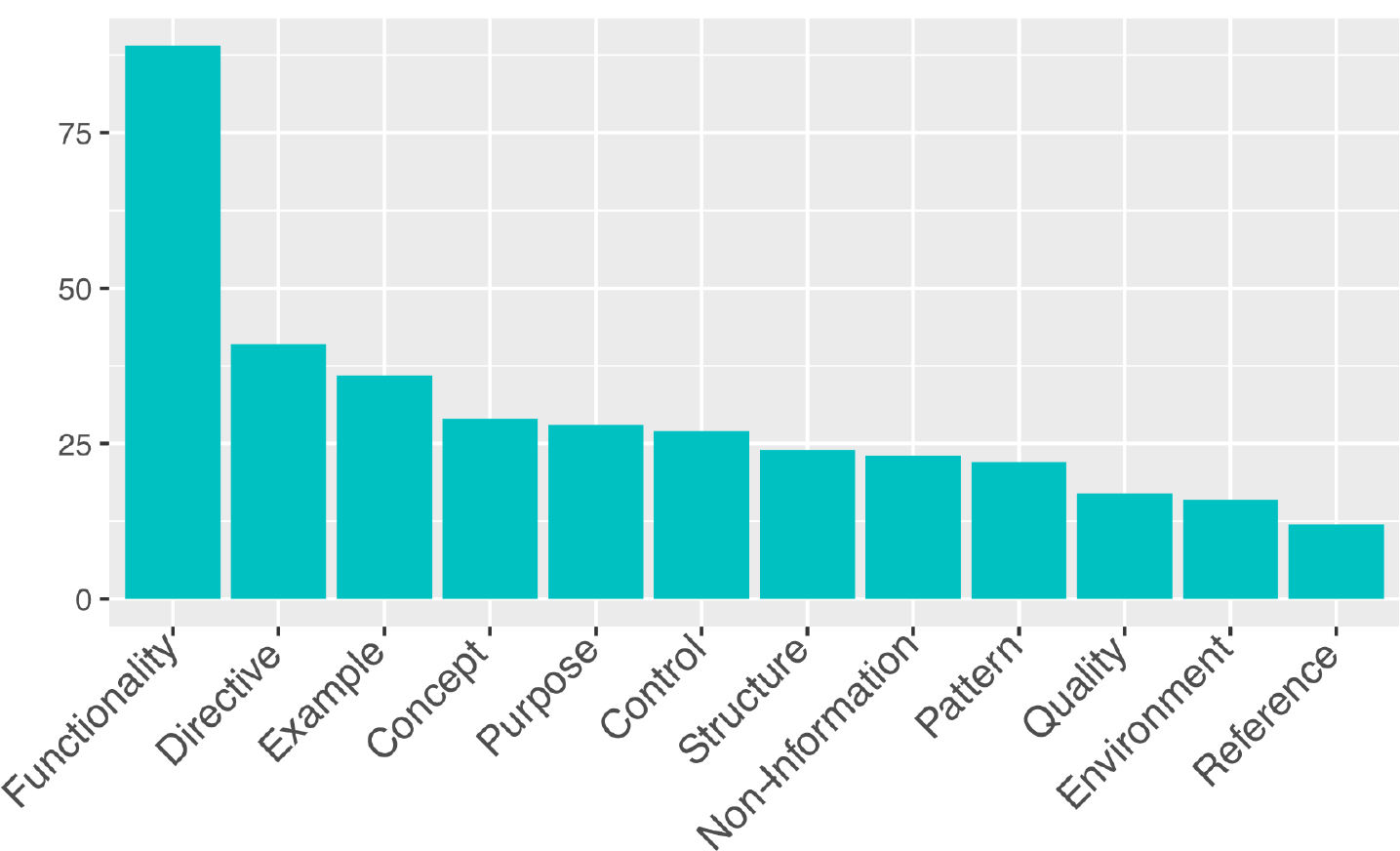}
    \caption{Knowledge types distribution in the \python dataset.}
    \label{fig:python}
\end{figure}

Two Ph.D. students in software engineering, accustomed to work with Python, manually labelled the knowledge types in each document.
For this task, we provided them the same guidelines from Maalej and Robillard\footnote{https://cado.informatik.uni-hamburg.de} with small adaptations, such as providing examples using the Python programming language.
The agreement on the label set was 14\%---i.e., 14 out of the 100 examples were labeled with the \textit{exact same} set of knowledge types. 
The overall agreement was 75\%---i.e., of the 1200 labels (100 examples $\times$ 12 labels), 300 were conflicting. 
Two of the authors addressed the conflicts and created the final dataset. 
\Cref{fig:python} shows the distribution of knowledge types in the \python dataset. \textit{Functionality} is the majority label.
However, as \python represents an additional test set (i.e., no examples from this dataset are used to train the classifiers), we did not resample it to avoid biased results.

 \begin{table*}[h!]
        \centering
    \caption{Summary of the corpora used to train the GloVe embedding.}
    \scalebox{0.85}{
    \begin{tabular}{lllll}
    \toprule
    ID & Name & Corpus description & \#docs & \#vocabulary \\ 
    \midrule
    \rowcolor[HTML]{EFEFEF} \textsc{CC} & Common Crawl  & General purpose, high-quality text crawled from Internet pages & 2.2M  & $\sim$220.000\\
    \textsc{CCotf} & Common Crawl on-the-fly & Common Crawl where missing words are learned from  \cado & 2.2M & $\sim$220.000\\ 
    \rowcolor[HTML]{EFEFEF} \textsc{SO} & StackOverflow  & StackOverflow questions and answers  & 20M & $\sim$400.000\\
    \textsc{SOapi} & StackOverflow Java and .NET posts  & StackOverflow questions and answers tagged as \texttt{java} or \texttt{.net} & 4M & $\sim$100.000\\ \bottomrule
    \end{tabular}
    }
    \label{tbl:1}
\end{table*}

For both datasets, we performed several, simple operations to clean and prepare the textual data. 
We lower-cased, tokenized, and applied stop-words removal to the API documentation text.
Then, we transformed terms in an order-preserving one-hot vectors.

% \textbf{Input Corpora.}
For the deep learning classifiers in our benchmark, we train GloVe~\cite{PSM14} embeddings based on four large corpora, summarized in~\Cref{tbl:1}.
The Common Crawl (\textsc{CC}) is a pre-trained embedding downloaded in March 2018.\footnote{\url{https://commoncrawl.org}}
It includes 840B tokens and a vocabulary of 2.2M words.
The corpus contains high-quality, general-purpose text crawled from the Internet.
However, the \textsc{CC} corpus is missing domain-specific terms present in the \cado dataset.  
Accordingly, in the \textsc{CCotf} embeddings, the missing words from the \textsc{CC} corpus are trained on-the-fly~\cite{HDG12}.
Finally, we obtained a completely domain-specific representation of the input by training embeddings on two additional corpora, StackOverflow (\textsc{SO}) and StackOverflow API (\textsc{SOapi}).
The former includes 20 million posts, while the latter includes 4 million posts tagged as \texttt{java} or \texttt{.net}. 

\section{Classifiers Configuration}
\label{sec:configs}
This section reports information about the configuration of both machine learning and deep learning classifiers used in this study.

\subsection{Traditional Machine Learning}
The machine learning approaches we selected for our classification task are SVM~\cite{Joa90} and \textit{k}-NN~\cite{BGR99} as well as their adaptations to multi-label problems, namely One-vs-Rest SVM (OvRSVM) and Multi Label \textit{k}-NN (ML-\textit{k}NN). 
We use unigrams and bigrams extracted from the \cado dataset as their input features as \textit{n}-gram language models are easy to compute and use.
Moreover, they have been used in studies where machine learning and natural language processing are applied to software engineering contexts~\cite{CK15,HBS12}. 

SVM is one of the most investigated approaches for statistical document classification and it is considered state-of-the-art~\cite{Joa06,Joa90}.
Moreover, it showed good results in software engineering-specific text classification problems (e.g.,~\cite{CLM18,FM17}).
SVM finds the hyper-plane maximizing the margin between two classes in the feature space, and it can learn and generalize high-dimensional features typical for text classification tasks~\cite{Joa90, Joa06}. 
When taking into account multiple knowledge types at once, we trained and reported the results of a SVM model adapted to such problem---i.e., OvRSVM using binary relevance~\cite{LZ05}. 
OvRSVM considers additional parameters and constraints necessary to solve the optimization problem with several classes and to handle the separation of several hyper-planes~\cite{LZ05}. 
For the SVM classifiers, we report the best model after hyper-parameters tuning using GridSearch~\cite{BB12}.

\textit{k}-NN is a widely-used approach in machine learning~\cite{CHP67}.
It determines the \textit{k} nearest neighbors of a document using Euclidean distance.
Then it assigns the document a label based on the document neighbors using Bayes decision rules~\cite{CHP67}. 
For the multi-label classification, we use ML-\textit{k}NN, which outperforms well-established multi-label classifiers~\cite{ZZ07}.

\subsection{RNN with LSTM layer}
Deep learning has recently brought substantial improvements in the field of machine vision and natural language processing (NLP)~\cite{DY14}.
The \lstm layer extends \rnn capabilities by utilizing several gates and a memory cell in the recurrent module to alleviate the vanishing gradient problem and to handle more efficiently the long-term dependencies between features~\cite{HS17}.  

\begin{figure}[!h]
\centering
    \includegraphics[width=0.5\textwidth]{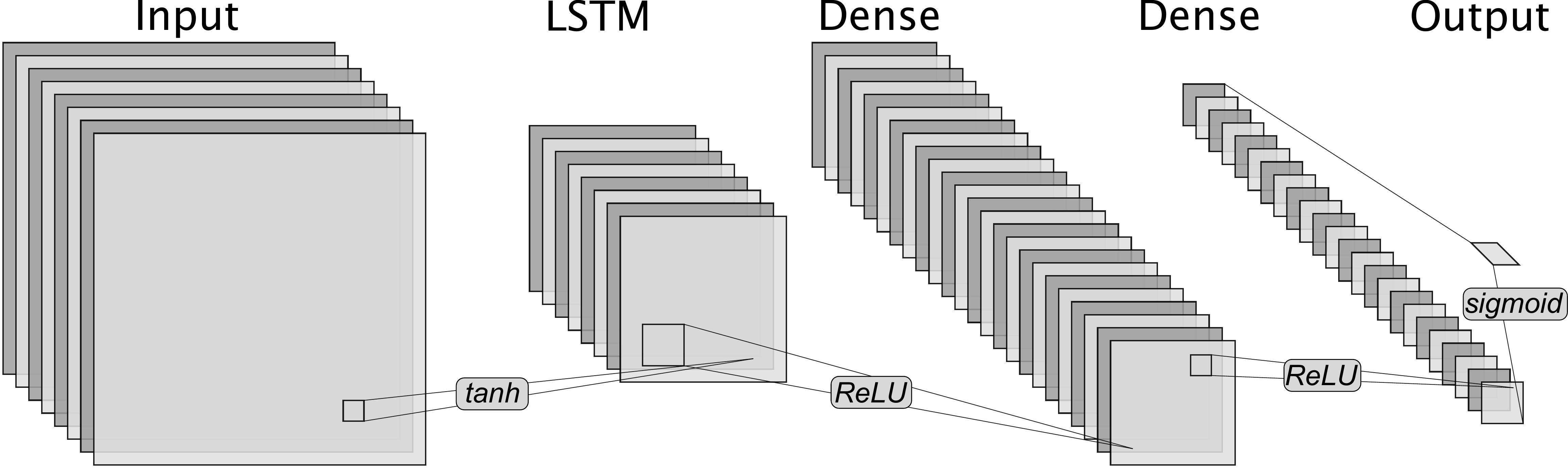}
    \caption{Architecture of the RNN used for classification of the knowledge types.}
    \label{fig:rnn}
\end{figure}
~\Cref{fig:rnn} shows the architecture we used in this work.
The network is composed of a single \lstm layer, two dense layers, and an output layer. 
The number of units in the \lstm layer is proportional to dimensions of the vectors used to represent each word. 
The dense layers contain 128 and 64 units respectively. 
The number of units in the output layer is the number of knowledge types (i.e., 12 units).

\begin{figure}[h]
    \centering
    \includegraphics[width=.35\textwidth]{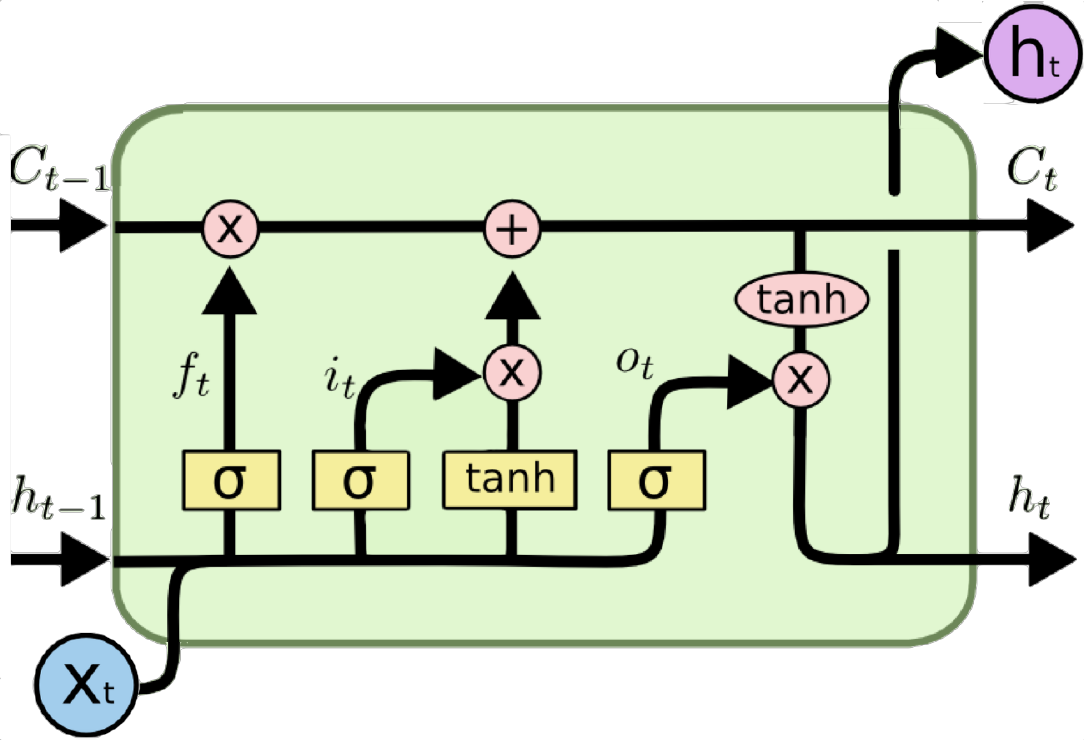}
    \caption{A single \lstm recurrent module containing input ($i_t$), output ($o_t$), and forgets gates ($f_t$).}
    \label{fig:memory}
\end{figure}

% \textbf{Long Short-Term Memory Networks.}
The core component of the \lstm layer is a memory cell which stores the information related to the previous analysis steps within the network. 
At each step of the training, the network predicts the output based on a) the new input, b) the previous state of the other hidden layers of the \rnn, and c) and the current state of the memory cell. 
Accordingly, the role of the gates is to learn how to modify the memory cell to enhance prediction accuracy (see~\Cref{fig:memory}).

The \textit{forget gate} ($\mathit{f_t}$) processes the information from the previous hidden state layer ($\mathit{h_{t-1}}$) and the current input ($X_t)$---i.e., a representation of the API documentation text.
It then decides what information should be discarded or kept from the previous state of the memory cell ($\mathit{C_{t-1}}$). 
% For example, if features associated with a new type are found (e.g., source code which is associated with the \textit{Example} type), the forget gate discards information (e.g., keywords) associated with the old knowledge type. 
The \textit{input gate} ($\mathit{i_t}$) is responsible for selecting new information from the input ($\mathit{X_t}$) that should be stored in the cell state. 
The third gate is called \textit{output gate} ($\mathit{o_t}$) and decides which part of the available information in the memory cell should be used to produce the final output ($\mathit{h_t}$).

The role of the forget cell in our network is to optimally discard information related to previous knowledge types when they change in the new input. 
For instance, when the knowledge type in the new input is \textit{Directive}, the forget gate removes the pieces of information associated with other knowledge types.
Consequently, the forget gate reduces the ambiguity of the memory cell when learning individual types. 
The features associated with the knowledge type in the current input document are moved into the memory cell.
In the memory cell, the input gate decides what information should be stored. 
For example, when the current input contains \textit{Directive}, its features will be extracted and stored in the memory cell using the input gate. 
Finally, the output gate selects the most significant features associated with the \textit{Directive} type. 

The \textbf{input layer} for the \rnn consists of word embedding vectors trained using GloVe~\cite{PSM14}.
GLoVe relies on the global occurrences of a word in a corpus by defining a word-to-word co-occurrence matrix. 
Each value of the matrix contains the probability $P$ of word $j$ appearing in the context of word $i$, as reported in Equation~\ref{formula:glove}.
In particular, $X_{ij}$ denotes the number of times word $j$ occurs in the context of word $i$ and $X_i$ denotes the number of times that any word $k$ appears in the context of word $i$.
Namely, GLoVe defines a learning function that estimates the probability ratio of the co-occurrence of two target words, $i$ and $j$, given a context word $k$~\cite{PSM14}.
\begin{equation}    
\label{formula:glove}
    P_{ij} = P(j|i) = X_{ij} / {X_i}, \; X_{i} = \sum_{k}X_{ik}.
\end{equation}

% Consequently, Pennington et al. (2014) shows that for the context word k related to i, not to j, the expected ratio of P ik /P jk will be rather high or vice versa. Moreover, if k is either related to both i and j, or to neither of them, the expected ratio should be close to one. Hence, it is straightforward to distinguish the related or not related words in the corpus. 
As input vectors with identical length speed up the process of building the embedding layer of the RNN~\cite{PG17}, we considered 300 as the maximum vector length and padded shorter vectors with zeroes. 
Therefore, our input layer is a 2D matrix where each row is a 300-dimensional unit (artificial neuron). 
The number of units in the input layer depends on the vocabulary size of the corpus used to train the embeddings. 
When a document is fed to the RNN, the units associated with the document terms will be activated.

The first \textbf{hidden layer} is a \lstm unit which learns textual features of an input document. 
It is suited to learn long-term dependencies, such as in the case of the large API reference documentation text.
To prevent over-fitting, we applied a dropout technique to the weights matrix and to the bias vector~\cite{SHK14}. 

The output of the \lstm layer (i.e., a set of features that can be associated with a knowledge type) goes through two fully-connected dense layers. 
The dense layers provide deep representations of the features extracted by the \lstm layer and enable the network to learn their hierarchical and compositional characteristics. 
To alleviate feature loss due to the projection of the features from a high-dimensional space to the low-dimensional space of the output layer, the model smoothly reduces the number of units in the dense layers from 128 to 64~\cite{DY14}. 
We use the ReLU activation function for the dense layers to prevent over-fitting~\cite{DY14}.

The \textbf{output layer} provides the predicted knowledge types using a sigmoid activation function. 
Hence, the number of units in the output layer is the number of labels that the model learns. 
As the output of the sigmoid is a probability value between 0 and 1, each neuron in this layer learns to estimate the probability of observing one of the labels. 
To binarize the predicted probabilities we used different thresholds according to the different metrics.
%For metrics such as precision and recall we set a threshold of 0.5, whereas for AUC and AUPRC we
% To calculate threshold-dependent metrics, such as AUC 

We tuned the following network parameters.\\
\textit{Epoch} is a complete pass (back and forward) of every sample through the neural network. As customary, we run 100 epochs~\cite{DY14}.\\
\textit{Batch size} is the number of samples passed through the network at once. As customary, we used a batch size of 32~\cite{DY14}.\\
\textit{Optimizer} is the optimization method minimizing the prediction error. We use Adam, a state-of-the-art algorithm for training RNN~\cite{Rud16}.\\
\textit{Loss Function} is the measure of the network prediction error. We use sigmoidal cross-entropy as it is efficient for text classification~\cite{NKM14}.\\
\textit{Learning Rate} is the parameter controlling the adjustments to the weights  with respect to the prediction error. We used a customary learning rate of .001~\cite{DY14}.

\begin{table*}[!t]
  \centering
  \caption{Comparison between Deep learning classifiers (trained with embeddings from general purpose and software development corpora), traditional machine learning, and na\"ive approaches for classifying \textit{individual} knowledge types in the \cado dataset. Values report the Area Under Precision-Recall Curve (AUPRC).}
% \usepackage{booktabs}
% \usepackage{multirow}
\scalebox{0.85}{
\begin{tabular}{lccccccccc}
\toprule
\multicolumn{1}{c}{} & \multicolumn{3}{c}{Na\"ive baselines} & \multicolumn{2}{c}{Traditional approaches} & \multicolumn{2}{c}{Deep learning (General Purpose)}  & \multicolumn{2}{c}{Deep learning (Software dev.)}
\\ \cmidrule(l){2-4} \cmidrule(l){5-6}\cmidrule(l){7-8}\cmidrule(l){9-10} 
\multicolumn{1}{c}{\multirow{-2}{*}{Knowledge Type}} & \multicolumn{1}{c}{\MFONE} & \multicolumn{1}{c}{\MFTWO} & \multicolumn{1}{c}{RAND} & \multicolumn{1}{c}{\textit{k}-NN} & \multicolumn{1}{c}{SVM} & \multicolumn{1}{c}{\CC} & \multicolumn{1}{c}{\CCOTF} & \multicolumn{1}{c}{\SO} & \multicolumn{1}{c}{\SOAPI} \\ \midrule
Functionality   & 0.69 & 0.73 & 0.72 & 0.76 & 0.39  & 0.86 & 0.84 & \textbf{0.87} & \textbf{0.87}   \\
\rowcolor[HTML]{EFEFEF} 
Concept        & 0.11 & 0.14 & 0.12 & 0.13 & \textbf{0.57}  & 0.25 & 0.28 & 0.28 & 0.28   \\
Directive      & 0.26 & 0.16 & 0.17 & 0.22 & 0.04  & 0.40 & 0.41 & 0.41 & \textbf{0.45}   \\
\rowcolor[HTML]{EFEFEF} 
Purpose         & 0.22 & 0.21 & 0.17 & 0.17 & 0.09  & 0.36 & 0.40 & 0.40 & \textbf{0.41}   \\
Quality         & 0.04 & 0.04 & 0.05 & 0.12 & 0.13  & \textbf{0.78} & 0.69 & 0.68 & 0.54   \\
\rowcolor[HTML]{EFEFEF} 
Control         & 0.08 & 0.12 & 0.09 & 0.08 & \textbf{0.81}  & 0.28 & 0.32 & 0.30 & 0.30   \\
Structure       & 0.37 & 0.37 & 0.35 & 0.38 & 0.42  & 0.61 & 0.56 & \textbf{0.63} & 0.60   \\
\rowcolor[HTML]{EFEFEF} 
Pattern       & 0.14 & 0.17 & 0.14 & 0.21 & \textbf{0.59}  & 0.46 & 0.46 & 0.48 & 0.51   \\
Example        & 0.24 & 0.23 & 0.20 & 0.25 & 0.60  & \textbf{0.90} & 0.85 & \textbf{0.90} & \textbf{0.90}   \\
\rowcolor[HTML]{EFEFEF} 
Environment     & 0.04 & 0.03 & 0.06 & 0.16 & 0.43  & 0.68 & \textbf{0.80} & 0.66 & 0.51   \\
Reference       & 0.11 & 0.14 & 0.16 & 0.13 & 0.15  & 0.35 & 0.35 & \textbf{0.41} & 0.30   \\
\rowcolor[HTML]{EFEFEF} 
Non-information & 0.29 & 0.31 & 0.28 & 0.33 & \textbf{0.71}  & 0.57 & 0.58 & 0.62 & 0.55  \\
\bottomrule
\end{tabular}
}
  \label{tbl:knowledge_types}
  \end{table*}
\section{Results}
\label{sec:results}
In this section, we compare the performance of RNN and traditional machine learning approaches for the classification of API reference documentation. 
We contrast the performance of RNN classifiers trained using different embeddings. 
Moreover, we assess the classifiers generalizability to another test set. 
% We use AUPRC to compare the two approaches; this metric is appropriate for problems that have unbalanced observations for each class. 
\subsection{Knowledge Types Identification}
\label{sec:rq.1}
\textbf{Individual knowledge types.}
We trained two RNN-based classifiers using a general-purpose corpus to create the embeddings for the input layer---\CC and \CCOTF.
~\Cref{tbl:knowledge_types} reports the evaluation of our classifiers.
RNN and traditional machine learning approaches improve over the na{\"i}ve baselines for all the individual knowledge type classification by up to 74\% (41\% on average).
SVM always performed better than \textit{k}-NN.  

Deep learning classify \textit{Functionality}, \textit{Example}, and \textit{Environment} with high precision and high recall at different probability thresholds (AUPRC $\geq$ 80\%) outperforming machine learning approaches.  
The RNNs yields subpar results for \textit{Directive}, \textit{Purpose}, \textit{Reference}, \textit{Concept}, and \textit{Control} (AUPRC $<$ 50\%). 
However, the best SVM outperforms the best RNN only for the latter two types. 

The best classifiers for \textit{Quality}, \textit{Structure}, \textit{Patterns}, and \textit{Non-information} yield an AUPRC between 59\% and 78\%. 
Also in this case, the best machine learning approach (i.e., SVM) outperforms the best RNNs classifiers only for the latter two types.
Compared to machine learning, RNNs classify better eight  knowledge types. 

\textbf{Multiple knowledge types.}
The second step is to consider our task as a multi-label classification problem rather than building individual classifiers for each knowledge type.
We compare the RNN classifiers to the multi-label adaptation of the same two machine learning models and two na{\"i}ve baselines (see \Cref{tbl:rq1}).  
% \KNN~\cite{ZZ07} is frequently used as a baseline into multi-label classification problems~\cite{HCR16}, whereas we implemented SVM as a One-versus-Rest (OvRSVM) classifier using binary relevance~\cite{RPH11} for transforming the problem in multi-label classification.  

ML-\textit{k}NN and OvRSVM perform worse than the baselines for the item-based metrics, whereas the RNNs shows the best performance.
The RNNs outperform ML-\textit{k}NN, OvRSVM, and the baselines for label-based metrics. 
There is an 11\% improvement regarding the most strict metric (i.e., Subset Accuracy) between the best RNN and machine learning classifiers.
Regarding MacroPrecision, MacroRecall, and MacroF1, there is an improvement between 25\% and 28\% for the RNN. 
MF1 performs better than traditional machine learning regarding MacroAUC, which RNNs improves by 17\%. 
% \CCOTF shows good results for both Hamming Loss and Subset Accuracy and improves over \CC
% Learning the embeddings of missing word on-the-fly (i.e.,~\CCOTF) improves item-based metrics as well MacroPrecision (10\%) and MacroAUC (1\%) with respect to \CC. 
% On the other hand, \CC shows better MacroRecall and MacroF1.  
% A two-sample Welch's t-test between \CC and \CCOTF ($p < .000$) confirms the results.\footnote{Both samples are normally distributed according to the results of the Anderson-Darling test.}

% As shown in~\Cref{tbl:rq1}, there are no substantial differences between the two classifiers using domain-specific embeddings,~\SO and \SOAPI.
% They perform worse than \CC and \CCOTF regarding item-based metrics, with a 25\% difference in terms of MacroRecall between \CC and \SO.
% Domain-specific embeddings increase the predictive power of the classifiers at most by 5\%, with \SOAPI yielding the best AUC. 
% The result of two-sample Welch's t-test confirms a significant difference in MacroAUC between \CC and \SOAPI ($p = 0.002$) and between \CCOTF and \SOAPI ($p = 0.001$).
% There is no significant difference between \CC and \SO ($p = 0.075$) and between \CCOTF and \SO ($p = 0.085$). 

\begin{mdframed}[backgroundcolor=lightgray!20,skipabove=5pt,skipbelow=0pt]
\textbf{Answer to RQ1.} One-third of the knowledge types can be automatically identified with good results (i.e., AUPRC $\ge$ 80\%). RNN can more accurately (> 10\%) identify eight of the 12 knowledge type compared to traditional machine learning approaches. When considering multi-label classification, RNN outperforms traditional machine learning approaches for item- and label-based metrics.
\end{mdframed}
\begin{table*}[!t]
\centering
\caption{Comparison between Deep learning classifiers (trained with embeddings from general purpose and software development corpora), traditional machine learning, and na\"ive approaches for classifying \textit{multiple} knowledge types in \cado.}
\scalebox{0.85}{
\begin{tabular}{lcccccccc}
\toprule
\multicolumn{1}{c}{}&
\multicolumn{2}{c}{Na{\"i}ve baselines}& 
\multicolumn{2}{c}{Traditional approaches}& 
\multicolumn{2}{c}{Deep learning (General Purpose)}& 
\multicolumn{2}{c}{Deep learning (Software dev.)}\\ 
\cmidrule(lr){2-3} 
\cmidrule(lr){4-5} 
\cmidrule(lr){6-7} 
\cmidrule(lr){8-9} 
\multirow{-2}{*}{Metric}&
\multicolumn{1}{c}{\model{MF1}} & 
\multicolumn{1}{c}{\model{MF2}} &
\multicolumn{1}{c}{\model{ML-\textit{k}NN}} & 
\model{OvRSVM}& 
\multicolumn{1}{c}{\model{RNN\textsubscript{CC}}} & 
\multicolumn{1}{c}{\model{RNN\textsubscript{\textsc{CCotf}}}} & 
\multicolumn{1}{c}{\model{RNN\textsubscript{SO}}} & 
\multicolumn{1}{c}{\model{RNN\textsubscript{\textsc{SOapi}}}}  
\\ 
\midrule
Hamming Loss & 0.17 &   0.20& 0.18 & 0.30 & 0.16 & \textbf{0.14} & \textbf{0.14} & \textbf{0.14} \\
\rowcolor[HTML]{EFEFEF}
Subset Accuracy  & 0.00 & 0.13& 0.11 & 0.02 & 0.20 & \textbf{0.22}  & 0.19 & 0.21 \\
\midrule
MacroPrecision & 0.05 &  0.08& 0.41 & 0.21 & 0.56 & \textbf{0.66} & 0.61 & 0.63 \\
\rowcolor[HTML]{EFEFEF}
MacroRecall & 0.16 & 0.16& 0.24 & 0.27& \textbf{0.55} & 0.39 & 0.30 & 0.33 \\
MacroF1 & 0.10 & 0.10& 0.27 & 0.24 & \textbf{0.55} & 0.44 & 0.40 & 0.43 \\
\rowcolor[HTML]{EFEFEF}
MacroAUC & 0.62 & 0.50& 0.55 & 0.61 & 0.73 & 0.74 & 0.78 & \textbf{0.79} \\

\bottomrule
\end{tabular}}
\label{tbl:rq1}
\end{table*}
% However, the results present a disparity between the MacroAUC and MacroF1.
% While MacroF1 does not account for true negatives, MacroAUC does as it takes into account the model specificity.
% The high true negatives rate is the result of the classifier correctly identifying knowledge types that are not in the document.
% True negatives can be inflated, for example, by the knowledge type appearing very infrequently in the test set.
% Based on this observation, we independently evaluated classifiers for each knowledge types.
\subsection{Software Development-specific Corpus}
\textbf{Individual knowledge types.}
One RNN uses freely-available, pre-trained embeddings based on a general purpose textual corpus (i.e., \CC), whereas \CCOTF uses the same corpus but learns missing words on-the-fly from the \cado dataset.
The assumption behind text statistical representations such as GloVe is that the meaning of a document is determined by the meaning of the words that appear in it~\cite{MC91}.
Accordingly, \SO and \SOAPI use corpora in a domain closer to the one of API documentation.

As shown in \Cref{tbl:knowledge_types}, the best among these RNNs performs similarly to their general domain counterparts ($\Delta$AUPRC = 4\%).
For \textit{Functionality}, \textit{Purpose}, \textit{Control}, and \textit{Structure} the differences are minimal (1-2\%). 
However, for \textit{Quality} and \textit{Environment} there is a substantial decrease in performance when using software development-specific embeddings (10\% and 14\%, respectively). 
Overall, the improvement is rather limited given the overhead in obtaining the corpus and computing the embeddings. 

\textbf{Multiple knowledge types.}
\Cref{tbl:rq1} shows that \CC and \CCOTF outperform \SO and \SOAPI when considering label-based metrics (except for MacroAUC) and perform similarly when considering item-based metrics.
The RNN trained using Java and .NET StackOverflow posts yields the best MacroAUC (79\%).

\begin{mdframed}[backgroundcolor=lightgray!20,skipabove=5pt,skipbelow=0pt]
\textbf{Answer to RQ2.} RNN using software development-specific embeddings show slight to no improvement over RNN using general purpose embeddings for classification of individual knowledge types. When considering multi-label learning, except for MacroAUC, using general purpose embeddings yields better results across item- and label-based metrics.
\end{mdframed}

\subsection{Classifiers Generalizability}
\label{sec:rq.2}
 \textbf{Individual knowledge types.}
~\Cref{tbl:knowledge_types_python} reports the performance of the individual RNN-based classifiers on the \python test set.
% Our approach yields a high AUPRC for the \textit{Functionality} knowledge type, with \SO scoring 0.95.
% The LSTMs using domain specific embeddings perform better than the general purpose ones for nine of the 12 knowledge type. 
Also in this setting, no na\"ive baseline performs better than traditional or deep learning approaches.
The RNNs are the best classifiers for seven knowledge types, whereas SVM shows the best results for the remaining five.
Consistently with the \cado setting, SVM is the best classifier for \textit{Concept}, \textit{Pattern}, and \textit{Non-information}.
Classifiers for \textit{Functionality}, \textit{Concept}, and \textit{Purpose} show some improved performance compared to the \cado settings ($\Delta$ AUPRC = 8.3\%).

There is a large absolute difference ($\Delta$ AUPRC = 33\%) between the two settings when considering \textit{Directive}, \textit{Quality}, \textit{Control}, \textit{Structure}, \textit{Example}, and \textit{Environment}, suggesting that these knowledge types are dependent on the settings. 
On average, the performance on the \python dataset decrease by $\sim$16\% over the 12 knowledge types.

\textbf{Multiple knowledge types.}
% Also in this setting, we studied the RNN performance classifying API documents based on multiple knowledge types.
\Cref{tbl:rq2} present the results of the multi-label classification task.
% Regarding item-based metrics, the LSTMs perform worse than the na\"ive baselines (see~\Cref{tbl:rq1}) 
Regarding item-based metrics, our classifiers perform worse or on par with respect to the na\"ive baselines. 
The classifiers show low precision (40\% for the best classifier, SVM) and recall (26\% for the best classifiers, \CCOTF and \SOAPI). SVM also achieves the best F1 (30\%).
% As in the case of the \cado dataset, there is a discrepancy between MacroF1 and MacroAUC. 
\SOAPI shows the best performance for MacroAUC (64\%) .

\begin{mdframed}[backgroundcolor=lightgray!20, skipabove=3pt,skipbelow=0pt]
\textbf{Answer to RQ3.} Classifiers for \textit{Functionality}, \textit{Concept}, \textit{Purpose}, \textit{Pattern}, and \textit{Directive} seem to generalize from Java and .NET to Python documentation. The generalization for multiple knowledge types classifiers is limited.
\end{mdframed}
% \usepackage{booktabs}
% \usepackage{multirow}
\begin{table*}[]
\centering
\caption{Comparison between Deep learning classifiers (trained with embeddings from general purpose and domain specific corpora), traditional machine learning, and na\"ive approaches for classifying API documents based on \textit{individual} knowledge type in the \python dataset. Values report the Area Under Precision-Recall Curve (AUPRC).}
\scalebox{0.85}{
\begin{tabular}{lccccccccc}
\toprule
\multicolumn{1}{c}{}&
\multicolumn{3}{c}{\begin{tabular}[c]{@{}l@{}}Na\"ive baselines\end{tabular}}& 
\multicolumn{2}{c}{\begin{tabular}[c]{@{}l@{}}Traditional approaches\end{tabular}}& 
\multicolumn{2}{c}{\begin{tabular}[c]{@{}l@{}}Deep learning (General Purpose)\end{tabular}}& 
\multicolumn{2}{c}{\begin{tabular}[c]{@{}l@{}}Deep learning (Software dev.)\end{tabular}}\\ 
\cmidrule(lr){2-4} 
\cmidrule(lr){5-6} 
\cmidrule(lr){7-8} 
\cmidrule(lr){9-10} 
\multirow{-2}{*}{Knowledge Type}&
\multicolumn{1}{c}{\model{MF1}} & 
\multicolumn{1}{c}{\model{MF2}} & 
\multicolumn{1}{c}{\model{RAND}} & 
\multicolumn{1}{c}{\model{\textit{k}-NN}} & 
\multicolumn{1}{c}{\model{SVM}} & 
\multicolumn{1}{c}{\model{RNN\textsubscript{CC}}} & 
\multicolumn{1}{c}{\model{RNN\textsubscript{\textsc{CCotf}}}} & 
\multicolumn{1}{c}{\model{RNN\textsubscript{SO}}} & 
\multicolumn{1}{c}{\model{RNN\textsubscript{\textsc{SOapi}}}}  \\
\midrule
Functionality & 0.89 & 0.89 & 0.92 & 0.85& 0.94 & 0.90 & 0.89 & \textbf{0.95} & 0.94\\
\rowcolor[HTML]{EFEFEF} 
Concept   & 0.29& 0.28& 0.31& 0.26& \textbf{0.64} &0.40 & 0.33& 0.49& 0.41\\
Directive & 0.41 & 0.41& 0.49& 0.42& \textbf{0.71} & 0.49& 0.44& 0.55& 0.63\\
\rowcolor[HTML]{EFEFEF} 
Purpose & 0.28& 0.28& 0.25&0.30 & 0.13 & 0.46& 0.40& \textbf{0.51}& 0.39\\
Quality & 0.17& 0.17&0.19 &0.17 & 0.27 & 0.20& 0.17& 0.20& \textbf{0.32}\\
\rowcolor[HTML]{EFEFEF} 
Control& 0.27& 0.27& 0.32&0.24 & 0.33 & 0.43& \textbf{0.46}& 0.39& 0.35\\
Structure &0.24 &0.24 &0.24 &0.32 &0.11 & 0.26& 0.24& 0.30& \textbf{0.32}\\
\rowcolor[HTML]{EFEFEF} 
Pattern& 0.22 &0.22 &0.24 &0.29 &\textbf{0.61} & 0.50& 0.30 & 0.41& 0.43\\
Example & 0.36 &0.36 &0.38 & 0.43&0.44 & 0.48& 0.49& \textbf{0.51}& 0.48\\
\rowcolor[HTML]{EFEFEF} 
Environment & 0.16 &0.16 &0.17 &0.16&\textbf{0.37}& 0.15& 0.15& 0.18& 0.17\\
Reference & 0.12  & 0.12& 0.17& 0.11&0.22 & 0.16 & 0.19 & 0.24 & \textbf{0.25} \\
\rowcolor[HTML]{EFEFEF} 
Non-information &0.23 & 0.23 &0.24 &0.27 &\textbf{0.61}& 0.30& 0.39& 0.30& 0.28\\ 
\bottomrule
\end{tabular}}
\label{tbl:knowledge_types_python}
\end{table*}
\begin{table*}[]
        \centering
        \caption{Comparison between Deep learning classifiers (trained with embeddings from general purpose and software development corpora), traditional machine learning, and na\"ive approaches for classifying \textit{multiple} knowledge types in \python.}
        \scalebox{0.85}{
        \begin{tabular}{lcccccccc}
        \toprule
\multicolumn{1}{c}{}&
\multicolumn{2}{c}{\begin{tabular}[c]{@{}l@{}}Na\"ive\end{tabular}}& 
\multicolumn{2}{c}{\begin{tabular}[c]{@{}l@{}}Traditional approaches\end{tabular}}& 
\multicolumn{2}{c}{\begin{tabular}[c]{@{}l@{}}Deep learning (General Purpose)\end{tabular}}& 
\multicolumn{2}{c}{\begin{tabular}[c]{@{}l@{}}Deep learning (Software dev.)\end{tabular}}\\ 
\cmidrule(lr){2-3} 
\cmidrule(lr){4-5} 
\cmidrule(lr){6-7} 
\cmidrule(lr){8-9} 
\multirow{-2}{*}{Metric}&
\multicolumn{1}{c}{\model{MF1}} & 
\multicolumn{1}{c}{\model{MF2}} & 
\multicolumn{1}{c}{\model{ML\textit{k}NN}} & 
\multicolumn{1}{c}{\model{OvRSVM}} & 
\multicolumn{1}{c}{\model{RNN\textsubscript{CC}}} & 
\multicolumn{1}{c}{\model{RNN\textsubscript{\textsc{CCotf}}}} & 
\multicolumn{1}{c}{\model{RNN\textsubscript{SO}}} & 
\multicolumn{1}{c}{\model{RNN\textsubscript{\textsc{SOapi}}}}  \\
\midrule
        Hamming Loss               & 0.23 & \textbf{0.25} &  0.28& 0.35 & 0.27                  & 0.30                     & 0.26                  & 0.27                     \\
        \rowcolor[HTML]{EFEFEF} 
        Subset Accuracy          &\textbf{0.05}&\textbf{0.05}& 0.02 & 0.01  & 0.02                  & 0.03                     & 0.04                  & \textbf{0.05}                     \\ \midrule
        MacroPrecision        &0.07&0.10& 0.33 &  \textbf{0.40}  & 0.36                  & 0.31                     & 0.31                  & 0.31                     \\
        \rowcolor[HTML]{EFEFEF} 
        MacroRecall            &0.08& 0.16& 0.24 &   0.24 & 0.24                  & \textbf{0.26}                     & 0.21                  & \textbf{0.26}                     \\
        MacroF1              &0.07&0.13&0.28&   \textbf{0.30}   & 0.29                  & 0.28                     & 0.25                  & 0.28                     \\
        \rowcolor[HTML]{EFEFEF} 
        MacroAUC            &0.50&0.50&0.53&  0.54     & 0.60                  & 0.57                     & 0.62                  & \textbf{0.64}                     \\ \bottomrule
        \end{tabular}}
        \label{tbl:rq2}
        \end{table*}
\section{Related Work}
\label{sec:related}
To the best of our knowledge, this is the first study, addressing the automated identification of \textit{several} knowledge types within API reference documentation.
In this section, we report related work investigating some of the knowledge types individually.
We present studies comparing traditional machine learning and deep learning approaches for text classification in software engineering.

\subsection{Knowledge Types in API Documentation}
% The systematic mapping study by Zhi \etal~\cite{ZGS15} summarizes the existing body of knowledge regarding API reference documentation collected from 69 papers published until 2011.
% Their results show that documentation quality and the methods to assess it are the main focus.
% In particular, the main quality attributes investigated in the literature are completeness, accuracy, and similarity.
% Our work can be leveraged to assess the quality of API reference documentation.
% For example, completeness refers to a clear contract specifying what the developer is allowed (or forbidden) to do with an API and implies the presence of the \textit{Directive} knowledge type. 
% Accuracy refers to the effort required to extract an use the information available in a reference document, usually supported by code examples, identified by \textit{Example} knowledge type.
% Similarity refers to avoiding content duplication which adds unnecessary cognitive effort when reading the documentation. 
% It can be identified by the presence of the \textit{Non-information} knowledge type.

Identifying a document based on the knowledge types it contains can support documentation quality assessment and improvement.
For example, Ding et al.~\cite{DLT14} systematic review of 60 primary studies investigates documentation quality attributes.
The authors focus on knowledge-based approaches used to address quality issues of API documentation.
Although retrievability is reported as an essential quality attribute, the authors show a lack of advanced ways to retrieve specific information from API documentation. 
On the one hand, our work represents a first step towards developing retrieval mechanisms for documents containing a set of knowledge types from the Java and .NET API reference documentation.
On the other hand, the individual classifiers showing a performance (e.g.,~\textit{Functionality}, \textit{Control}, \textit{Example}, and \textit{Environment}) can be used to retrieve documents containing a specific knowledge type.
Moreover, our classifiers can be used to retrieve documents containing \textit{Functionality} from the Python standard library documentation. 

Previous research tried to automatically retrieve particular knowledge from API documentation.
Robillard and Chhetri~\cite{RC14} presented an approach to identify API-related information that developers should not ignore as well as non-critical information.
Their approach---based on natural language analysis (i.e., part-of-speech tagging, word patterns)---shows 90\% precision and 69\% recall when applied to 1000 Java documentation units. 
However, the authors needed to manually assess, on top of the sensible knowledge items, also obvious, unsurprising, and predictable documentation---i.e., what we consider \textit{Non-information}~\cite{RC14}.
Our SVM classifier, trained using simple features, identifies \textit{Non-information} with 71\% accuracy.

Montperrus et al.~\cite{MET11} studied a particular knowledge type found in API reference documentation, \textit{Directive}. 
They analyzed more than 4000 API documentation from open source libraries.
To determine the documents containing \textit{Directive}, they developed a set of syntactic patterns associated with concerns reported in the documentation.
Finally, they manually created a taxonomy of 23 directives.
% On top of reporting how frequent each type of directives are (65\% of the API contains at least one of the directive in the taxonomy), the authors propose guidelines to write directive documentation. 
% Similarly, Saied \etal~\cite{SSD15} studied four usage constraints reported in the Java API documentation.
% Such constraints are associated with the \textit{Directive} knowledge type.
% The analysis of 11 large Java API projects showed that undocumented usage constraints are approximately 80\% more frequent than documented ones.
Pandita et al.~\cite{PXZ12} proposed an NLP-based approach to verify the legal usage of API methods against its description extracted automatically from the documentation.
Their approach uses features derived from part-of-speech tagging and chunking techniques to semantically analyze text.
Moreover, using a domain dictionary, the authors extracted methods specifications as first-order logic expressions to verify their legal usage in client code.
% Studies similar to the ones presented above can benefit from a first automatic identification of \textit{Directive} within API documentation, as presented in this paper.

Conversely, in this work, we attempted a simple approach based only on features which can be automatically extracted from the raw text.
Our goal was to create a benchmark which can be improved by including, for example, natural language patterns specific for each knowledge types and domain-specific models.  
We show that some classifiers have already practical relevance.
% Saied \etal~\cite{SBA15} presented a method to identify API usage patterns.
% The authors propose to enhance the API documentation with information about usage patterns which is usually missing or difficult to comprehend. 
% Their proposed approach processes a program that uses an API and extracts the API methods using static analysis. 
% The methods co-usage is represented by a binary vector.  
% The co-usage patterns are extracted using a hierarchical clustering algorithm based on DBSCAN.
% Similar studies can benefit from automatically establishing a baseline of \textit{Patterns} already reported in API documentation using the approach proposed in this paper.

\subsection{Deep Learning in Software Engineering}
Xu et al.~\cite{XYX16} use CNN to semantically link together knowledge units from StackOverflow. 
Their approach focuses on predicting several classes of relatedness (e.g., duplicate, related information).
The network input is the word2vec representation of 100,000 Java-related posts from StackOverflow, whereas the dataset includes 8,000 knowledge units balanced among relatedness types.
The CNN outperformed machine learning baselines---i.e., SVM trained using tf-idf and word2vec.
However, Fu and Menzies~\cite{FM17} replicated Xu et al. study comparing their results to the same SVM baselines optimized using hyper-parameter tuning. 
The authors showed improved results for the baselines which perform closely (if not better) to the CNN, although the latter required 84x more time to train.
In this work, we also used a deep learning approach with a semantic representation of the input based on StackOverflow.
We found that, for our task, there are only a few small improvements due to the software development-specific corpus, which may not be worth when considering the extra effort required to obtain and train the embeddings. 
We compared the deep learning approach to (among others) SVM models trained in line with the suggestions of Fu and Menzies~\cite{FM17}. We showed that the approaches are complementary as their performance depends on the specific knowledge types.

Fakhoury \etal~\cite{FAV18} applied deep learning and traditional machine learning to the detection of language anti-patterns in software artifacts (e.g., poor naming conventions and documentation) using a dataset of 1,700 elements collected from 13 large Java system. 
The authors showed that using Bayesian optimization and model selection, traditional machine learning models can outperform deep learning not only in accuracy but also regarding the use of computational resources.
They advise researchers and practitioners to explore traditional machine learning models with hyper-parameter tuning before turning to deep learning approaches.
Our results for individual knowledge types partly support this conclusion. 
However, when tackling multi-label problems, our work shows that deep learning performs better than traditional machine learning for all the reported metrics. 
\section{Discussion}
\label{sec:discussion}
In this section, we discuss the implications for practitioners and researchers.
Then, we present the limitations of this study. 

\subsection{Implications}
\textbf{Building automated knowledge extraction tools.}
Classifiers showing good performance (AUPRC $\ge$ 80\%) can already be used in practice to tag documents containing crucial information for developers.
Moreover, these classifiers are trained using either traditional machine learning algorithms with simple text features or using deep learning but with readily available embeddings. 

A document containing \textit{Functionality} can answer developers' information needs regarding what the API does, whereas \textit{Control} and \textit{Example} address how to accomplish a task using the API. 
The classifier for \textit{Functionality} can be applied also to Python documentation. 
The \textit{Environment} classifiers can be used to get information regarding an API usage context.
Classifiers for \textit{Quality} and \textit{Non-information} showed encouraging results (AUPRC $\ge$ 70\%). 
The former is relevant to understand API performance, whereas the latter is useful for suggesting information that a developer can ignore.
Moreover, the \textit{Non-information} classifier showed promising results generalizing to the Python documentation.
Given its particular use case, we suggest research to focus on maximizing recall to ensure that \textit{all} uninformative documents can be tagged appropriately.
For the other knowledge types, we suggest maximizing precision to guarantee that fundamental information is correctly tagged.

The results for other knowledge types can be improved by adding NLP-based features. 
For example, \textit{Structure} usually contains references to other API elements that can be identified using a specific named-entity tagger (e.g.,~\cite{MCC18}).
\textit{Concept} and \textit{Pattern} are strongly characterized by explanations of specific terms and sequence of steps. 
These can be identified through specialized features based on linguistic inquiry~\cite{TP10}, such as \textit{drives} (e.g., ``do this to achieve that'') and \textit{time orientation} (e.g., ``do this, then do that'').  
As these classifiers showed similar results when applied to the Python documentation, their improvement can also increase their generalizability.

The classifiers showed the worst results (AUPRC $<$ 50\%) for \textit{Directive}, \textit{Purpose}, and \textit{Reference} knowledge types.
The first two can be the subject of further research. 
In particular, features for a \textit{Directive} classifier can be extracted from Maalej and Robillard work~\cite{MR13} as well as from the specific taxonomy developed by Montperrus et al.~\cite{MET11}.
Furthermore, previous work on rationale mining for other software engineering tasks (e.g.,~\cite{KM18, RGQ12}) can be adapted to improve the results for the \textit{Purpose} knowledge type. 
The \textit{Reference} classifier showed some of the weakest performance but it can be improved with simple syntactical features---e.g., the presence of links. 

% A first use leverage some of the classifiers presented in this work is the - First thing: add tags. Perhaps not all types. ~7? Non-Info useful when the only type in the docu (High recall, while high precision for the others critical types).
% Some of the classifiers can already be used. 
% Classifiers can be further tuned
% - Adding syntactical features
% - Types of API element
% - NLP pattern…
% Discuss results of binary classifiers one by one and of too bad (and no clear idea how to improve classifier, then recommend do not provide as a tag. More for research). 
% At least FOR JAVA/NET, API that follows these framework.
% Tags that worked for Py. can be generalized.
There is a variation in performance between the classifier configurations (e.g., traditional machine learning vs. deep learning) and between the individual knowledge types.
We hypothesize that some knowledge types can be sensitive to specific keywords, such as ``callback,'' ``event,'' and ``trigger'' in the case of \textit{Control}.
On the other hand, knowledge types such as \textit{Environment} and \textit{Example} are characterized by a change in the language context. 
The former tends to interpolate text with numbers (as it includes information such as version and copyright year), while the latter contains sequences that do not occur in natural language (i.e., source code). 
We postulate that the RNN can capture this change of context.
However, the explanations for some classifiers results are more subtle.
For example, \textit{Non-information} implies expressing in natural language information already provided by a method signature. 
This implies a mapping between source code tokens and natural language ones which need to be further investigated.
Similarly, the \textit{Purpose} knowledge type contains information---i.e., the answer to a ``why'' question---which can be difficult to identify, from a semantic perspective, using the simple configurations of our classifiers.
Arguably, the intrinsic difficulty to identify a knowledge type, even for a human expert, can explain some of the poor results.  
For instance, Maalej and Robillard report low agreement for \textit{Purpose}---a knowledge type showing subpar results (AUPRC = 41\%).

Another explanation for the different performance between traditional machine learning and deep learning can lay in the parameters used to tune the latter.
A suggested improvement is to create 12 binary RNNs (one for each knowledge type) and select different parameters for a) the activation function of the output layer (e.g., SoftMax~\cite{HS09}), b) the loss function (e.g., Categorical Cross-Entropy~\cite{ZS18}), and c) the optimizer (e.g., RMSProp or ADA~\cite{MH17}).

\textbf{Using knowledge types when developing software.} 
One of the main applications for the classifiers presented in this study is documentation filtering.
On top of the current option (e.g., by package or class), API websites can offer their users the possibility to search documentation based on specific knowledge types.
For example, a developer fixing a specific performance bug (e.g., related to wireless connectivity) can search the network API documentation containing the \textit{Quality} knowledge type. 
In this scenario, the classifier can be optimized for precision---i.e., the developer would consult a small number of documents which are likely to contain the information she needs.  
On the other hand, a developer exploring possible usage of a new set of APIs can filter them according to \textit{Functionality} which describes the their capabilities.
In this scenario, the classifier can be optimized for recall---i.e., the developer consults a substantial amount of documents which offer her a complete overview of the API functionalities, even if some may be irrelevant. 
Our benchmark is a starting point for selecting which classifier to optimize according to specific scenarios.
Classifiers with AUPRC $\ge$ 80\% can already be utilized the scenarios above. 
Classifiers with AUPRC $\ge$ 50\% need further optimization.

The proposed benchmark is also a stepping stone to support software developers filtering API documentation based on \textit{multiple} knowledge types of interest.  
Given the complexity of such a task, our best classifier (\SOAPI) showed good results (MacroAUC = 79\%).
However, when disregarding recall---based on the assumption that a developer will not read a large number of documents---the classifier with the highest precision (66\%) is \CCOTF. 
Conversely, when a developer can tolerate noisy yet comprehensive results, we recommend using \CC---i.e., the classifier with the best recall.
In both cases, the classifiers rely on ``affordable'' embeddings.

% Fixing a bug related to performance: might filter API doc related to quality. Discuss precision and recall for 1-2 scenarios (each one is needed in a different setting? Thus a benchmark is useful).
% Option to correct the model if developer notice a documentation is not labeled correctly.

% The proposed \rnn accurately identifies knowledge types which are fundamental for software developers (i.e., the best AUPRC for \textit{Functionality}, \textit{Examples}, and \textit{Environment} are 0.86, .90, and 0.80 respectively) without the need of training an ad-hoc embedding while reusing pre-trained ones. 
\textbf{Using knowledge types when authoring  documentation.}
API documentation providers can leverage the results of this work to monitor their product. 
For example, they can use simple machine learning models (e.g., SVM) to find documents containing \textit{Non-information} and remove irrelevant text that increase the developers' cognitive effort (e.g., the repetition of a method signature in textual form).
Furthermore, they can monitor the presence of knowledge types containing crucial information for software developers, such as \textit{Functionality}, \textit{Control}, and \textit{Example}.
API documentation provider can also monitor the decrease of important knowledge types (e.g., \textit{Functionality}) or increase of harmful ones (i.e., \textit{Non-information}) before releasing new version of an API and its documentation. 
API defects can be diagnosed by identifying (and subsequently improving) documentation containing \textit{Directive} and \textit{Quality}.   

% \textbf{For practitioners.}

% However, for practical use, both tasks require a better F1-measure than the current 0.55 provided by our best classifier (\CC).

\textbf{Further research outlets.}
Researchers investigating documentation quality can benefit from the results of our work.
For example, quality models can be devised based on the presence (or absence) of specific types.
A first step is the identification of knowledge types in a set of documents.
In our benchmark, the \SOAPI model showed good results (MacroAUC = 0.79). 
The classifier correctly identifies documents containing a set of knowledge types with 60\% false positives rate when maximizing recall. 

Researchers should also consider the trade-off between using a pre-trained embedding while losing some performance (5\%) in terms of MacroAUC. 
Given the results obtained on the Python standard library, we recommend researchers to be careful when applying our multi-label models to different API documentation. 
Researchers have shown interest in studying how the usage of particular elements in a framework is documented (\eg~\cite{{RC14,SSD15,MET11}}).
This line of research can benefit from an approach to automatically retrieve API reference documentation containing the \textit{Functionality} knowledge type using the \SO model, as it showed good performance on different test sets.

\subsection{Threats to Validity}
The API reference documentation used to train our classifiers is based on two libraries, JDK and .NET.
While the language paradigms are similar, their documentation styles are different~\cite{MR13}. 
Moreover, we directly addressed a threat to generalizability by investigating the less structured  documentation of the Python programming language API~\cite{DR10}.
% On the other hand, the Python API reference documentation used to investigate RQ3 presents some differences---\eg it is less structured~\cite{DR10}.
We acknowledge that our results may not hold for API reference documentation in other domains (e.g., for a specific framework) or for a different programming paradigm (e.g., declarative programming).
Although Maalej and Robillard taxonomy is general enough~\cite{MR13}, other knowledge types may exist.

The labeling of our new test set can introduce a threat to internal validity.
To mitigate such threat, two raters independently labeled the documents using validated guidelines~\cite{MR13}.
We reconciled the disagreements (approximately 50\% were clear mistakes) by discussing borderline cases and reaching consensus among the authors.

Our benchmark only includes two traditional machine learning algorithms, one specific deep learning architecture, and four representations for the RNN input layer.
Nevertheless, there may be other algorithms, embeddings, and configurations worth of investigation. 

The results can be biased due to the unbalancedness of the dataset. 
To reduce this threat, we applied common resampling techniques to the training set and reported the performance according to appropriate metrics.
We did not observe a correlation between the classifiers performance and the distribution of the labels. 
% The \cado dataset contains enough occurrences for the single knowledge types (threat limited at least for the multiple binary)
\section{Conclusion and Future Work}
\label{sec:conclusion}
In this paper, we built several classifiers, using machine learning and deep learning approaches, to automatically identify the 12 knowledge types proposed by Maalej and Robillard~\cite{MR13} in API documentation. 
We used Java and .NET manually-annotated API documentation (n = 5,574) as a dataset for training and testing the classifiers.
We showed good results (i.e., AUPRC $\ge$ 80\%) for one-third of the knowledge types. 
In particular, RNN identifies eight types more accurately than traditional machine learning.
When considering multiple knowledge types at once (i.e., multi-label classification), RNN outperforms traditional machine learning approaches.

When word embeddings (i.e., the RNN input layer) are created from StackOverflow posts, rather than from general-purpose text, there is slight to no improvement in performance ($\Delta$AUPRC = 4\%). 
When considering multiple labels, software development-specific embeddings yield better results for MacroAUC (79\% vs. 74\%).

We applied the classifiers to a new test set (n = 100) obtained from the Python API documentation. 
Classifiers for \textit{Functionality}, \textit{Concept}, \textit{Purpose}, \textit{Pattern}, and \textit{Directive} generalize to Python. 
However, the generalization of multi-label classifiers is limited.

Some of the classifiers presented in this work can be already used by practitioners (e.g., developers, API providers) in different application scenarios.
Moreover, we propose possible improvements to the classifiers based on features specific for a knowledge type.

For further studies, we plan to evaluate other deep learning and hybrid approaches.
Based on our benchmark, we plan to implement a tool (e.g., a browser plugin) to filter API documentation based on knowledge types and evaluate its usefulness with software developers.
Finally, our long-term goal is to achieve knowledge identification in specific sentences or paragraphs of API documentation.

\bibliographystyle{IEEEtran}
\bibliography{main.bib} 

\begin{thebibliography}{10}
\providecommand{\url}[1]{#1}
\csname url@samestyle\endcsname
\providecommand{\newblock}{\relax}
\providecommand{\bibinfo}[2]{#2}
\providecommand{\BIBentrySTDinterwordspacing}{\spaceskip=0pt\relax}
\providecommand{\BIBentryALTinterwordstretchfactor}{4}
\providecommand{\BIBentryALTinterwordspacing}{\spaceskip=\fontdimen2\font plus
\BIBentryALTinterwordstretchfactor\fontdimen3\font minus
  \fontdimen4\font\relax}
\providecommand{\BIBforeignlanguage}[2]{{%
\expandafter\ifx\csname l@#1\endcsname\relax
\typeout{** WARNING: IEEEtran.bst: No hyphenation pattern has been}%
\typeout{** loaded for the language `#1'. Using the pattern for}%
\typeout{** the default language instead.}%
\else
\language=\csname l@#1\endcsname
\fi
#2}}
\providecommand{\BIBdecl}{\relax}
\BIBdecl

\bibitem{RD10}
M.~P. Robillard and R.~DeLine, ``{A field study of API learning obstacles},''
  \emph{Empirical Software Engineering}, vol.~16, no.~6, pp. 703--732, 2010.

\bibitem{DH09}
U.~Dekel and J.~D. Herbsleb, ``Improving api documentation usability with
  knowledge pushing,'' in \emph{Proceedings of the 31st International
  Conference on Software Engineering}.\hskip 1em plus 0.5em minus 0.4em\relax
  IEEE Computer Society, 2009, pp. 320--330.

\bibitem{MR13}
W.~Maalej and M.~P. Robillard, ``{Patterns of Knowledge in API Reference
  Documentation},'' \emph{IEEE Trans. Softw. Eng.}, vol.~39, no.~9, pp.
  1264--1282, 2013.

\bibitem{PRD15}
G.~Petrosyan, M.~P. Robillard, and R.~De~Mori, ``Discovering information
  explaining api types using text classification,'' in \emph{Proceedings of the
  37th International Conference on Software Engineering-Volume 1}.\hskip 1em
  plus 0.5em minus 0.4em\relax IEEE Press, 2015, pp. 869--879.

\bibitem{RC14}
M.~P. Robillard and Y.~B. Chhetri, ``{Recommending reference API
  documentation},'' \emph{Empirical Software Engineering}, vol.~20, no.~6, pp.
  1558--1586, Jul. 2014.

\bibitem{SGB08}
J.~Stylos, B.~Graf, D.~K. Busse, C.~Ziegler, R.~Ehret, and J.~Karstens, ``A
  case study of api redesign for improved usability,'' in \emph{Visual
  Languages and Human-Centric Computing, 2008. VL/HCC 2008. IEEE Symposium
  on}.\hskip 1em plus 0.5em minus 0.4em\relax IEEE, 2008, pp. 189--192.

\bibitem{SM08}
J.~Stylos and B.~A. Myers, ``The implications of method placement on api
  learnability,'' in \emph{Proceedings of the 16th ACM SIGSOFT International
  Symposium on Foundations of software engineering}.\hskip 1em plus 0.5em minus
  0.4em\relax ACM, 2008, pp. 105--112.

\bibitem{MET11}
M.~Monperrus, M.~Eichberg, E.~Tekes, and M.~Mezini, ``{What should developers
  be aware of? An empirical study on the directives of API documentation},''
  \emph{Empirical Software Engineering}, vol.~17, no.~6, pp. 703--737, 2011.

\bibitem{SSD15}
M.~A. Saied, H.~Sahraoui, and B.~Dufour, ``An observational study on api usage
  constraints and their documentation,'' in \emph{Software Analysis, Evolution
  and Reengineering (SANER), 2015 IEEE 22nd International Conference on}.\hskip
  1em plus 0.5em minus 0.4em\relax IEEE, 2015, pp. 33--42.

\bibitem{DY14}
L.~Deng and D.~Yu, ``Deep learning: methods and applications,''
  \emph{Foundations and Trends{\textregistered} in Signal Processing}, vol.~7,
  no. 3--4, pp. 197--387, 2014.

\bibitem{HS17}
S.~Hochreiter and J.~Schmidhuber, ``Long short-term memory,'' \emph{Neural
  computation}, vol.~9, no.~8, pp. 1735--1780, 1997.

\bibitem{LBH15}
Y.~LeCun, Y.~Bengio, and G.~Hinton, ``Deep learning,'' \emph{Nature}, vol. 521,
  no. 7553, p. 436, 2015.

\bibitem{MSC13}
T.~Mikolov, I.~Sutskever, K.~Chen, G.~S. Corrado, and J.~Dean, ``Distributed
  representations of words and phrases and their compositionality,'' in
  \emph{Advances in neural information processing systems}, 2013, pp.
  3111--3119.

\bibitem{BEP13}
K.~Boyd, K.~H. Eng, and C.~D. Page, ``Area under the precision-recall curve:
  Point estimates and confidence intervals,'' in \emph{Joint European
  Conference on Machine Learning and Knowledge Discovery in Databases}.\hskip
  1em plus 0.5em minus 0.4em\relax Springer, 2013, pp. 451--466.

\bibitem{HL05}
J.~Huang and C.~X. Ling, ``Using auc and accuracy in evaluating learning
  algorithms,'' \emph{IEEE Transactions on knowledge and Data Engineering},
  vol.~17, no.~3, pp. 299--310, 2005.

\bibitem{SL09}
M.~Sokolova and G.~Lapalme, ``A systematic analysis of performance measures for
  classification tasks,'' \emph{Information Processing \& Management}, vol.~45,
  no.~4, pp. 427--437, 2009.

\bibitem{CRD14}
F.~Charte, A.~Rivera, M.~J. del Jesus, and F.~Herrera, ``Concurrence among
  imbalanced labels and its influence on multilabel resampling algorithms,'' in
  \emph{International Conference on Hybrid Artificial Intelligence
  Systems}.\hskip 1em plus 0.5em minus 0.4em\relax Springer, 2014, pp.
  110--121.

\bibitem{HCR16}
F.~Herrera, F.~Charte, A.~J. Rivera, and M.~J. Del~Jesus, ``Multilabel
  classification,'' in \emph{Multilabel Classification}.\hskip 1em plus 0.5em
  minus 0.4em\relax Springer, 2016, pp. 17--31.

\bibitem{PSM14}
J.~Pennington, R.~Socher, and C.~D. Manning, ``Glove: Global vectors for word
  representation.'' in \emph{EMNLP}, vol.~14, 2014, pp. 1532--1543.

\bibitem{HDG12}
\BIBentryALTinterwordspacing
G.~Halawi, G.~Dror, E.~Gabrilovich, and Y.~Koren, ``Large-scale learning of
  word relatedness with constraints,'' in \emph{KDD}.\hskip 1em plus 0.5em
  minus 0.4em\relax New York, NY, USA: ACM, 2012, pp. 1406--1414. [Online].
  Available: \url{http://doi.acm.org/10.1145/2339530.2339751}
\BIBentrySTDinterwordspacing

\bibitem{Joa90}
T.~Joachims, ``Text categorization with support vector machines: Learning with
  many relevant features,'' in \emph{European conference on machine
  learning}.\hskip 1em plus 0.5em minus 0.4em\relax Springer, 1998, pp.
  137--142.

\bibitem{BGR99}
K.~Beyer, J.~Goldstein, R.~Ramakrishnan, and U.~Shaft, ``When is ``nearest
  neighbor'' meaningful?'' in \emph{International conference on database
  theory}.\hskip 1em plus 0.5em minus 0.4em\relax Springer, 1999, pp. 217--235.

\bibitem{CK15}
C.~A. Cois and R.~Kazman, ``Natural language processing to quantify security
  effort in the software development lifecycle.'' in \emph{SEKE}, 2015, pp.
  716--721.

\bibitem{HBS12}
A.~Hindle, E.~T. Barr, Z.~Su, M.~Gabel, and P.~Devanbu, ``On the naturalness of
  software,'' in \emph{2012 34th International Conference on Software
  Engineering (ICSE)}.\hskip 1em plus 0.5em minus 0.4em\relax IEEE, 2012, pp.
  837--847.

\bibitem{Joa06}
T.~Joachims, ``Training linear svms in linear time,'' in \emph{Proceedings of
  the 12th ACM SIGKDD international conference on Knowledge discovery and data
  mining}.\hskip 1em plus 0.5em minus 0.4em\relax ACM, 2006, pp. 217--226.

\bibitem{CLM18}
F.~Calefato, F.~Lanubile, F.~Maiorano, and N.~Novielli, ``Sentiment polarity
  detection for software development,'' \emph{Empirical Software Engineering},
  vol.~23, no.~3, pp. 1352--1382, 2018.

\bibitem{FM17}
W.~Fu and T.~Menzies, ``Easy over hard: A case study on deep learning,'' in
  \emph{Proceedings of the 2017 11th Joint Meeting on Foundations of Software
  Engineering}.\hskip 1em plus 0.5em minus 0.4em\relax ACM, 2017, pp. 49--60.

\bibitem{LZ05}
Y.~Liu and Y.~F. Zheng, ``One-against-all multi-class svm classification using
  reliability measures,'' in \emph{Proceedings. 2005 IEEE International Joint
  Conference on Neural Networks, 2005.}, vol.~2.\hskip 1em plus 0.5em minus
  0.4em\relax IEEE, 2005, pp. 849--854.

\bibitem{BB12}
J.~Bergstra and Y.~Bengio, ``Random search for hyper-parameter optimization,''
  \emph{Journal of Machine Learning Research}, vol.~13, no. Feb, pp. 281--305,
  2012.

\bibitem{CHP67}
T.~M. Cover, P.~E. Hart \emph{et~al.}, ``Nearest neighbor pattern
  classification,'' \emph{IEEE transactions on information theory}, vol.~13,
  no.~1, pp. 21--27, 1967.

\bibitem{ZZ07}
M.-L. Zhang and Z.-H. Zhou, ``Ml-knn: A lazy learning approach to multi-label
  learning,'' \emph{Pattern recognition}, vol.~40, no.~7, pp. 2038--2048, 2007.

\bibitem{PG17}
J.~Patterson and A.~Gibson, \emph{Deep Learning: A Practitioner's
  Approach}.\hskip 1em plus 0.5em minus 0.4em\relax O'Reilly Media, 2017.

\bibitem{SHK14}
\BIBentryALTinterwordspacing
N.~Srivastava, G.~E. Hinton, A.~Krizhevsky, I.~Sutskever, and R.~Salakhutdinov,
  ``Dropout: a simple way to prevent neural networks from overfitting.''
  \emph{Journal of Machine Learning Research}, vol.~15, no.~1, pp. 1929--1958,
  2014. [Online]. Available:
  \url{http://www.cs.toronto.edu/~rsalakhu/papers/srivastava14a.pdf}
\BIBentrySTDinterwordspacing

\bibitem{Rud16}
S.~Ruder, ``An overview of gradient descent optimization algorithms,''
  \emph{arXiv preprint arXiv:1609.04747}, 2016.

\bibitem{NKM14}
J.~Nam, J.~Kim, E.~L. Menc{\'\i}a, I.~Gurevych, and J.~F{\"u}rnkranz,
  ``Large-scale multi-label text classification---revisiting neural networks,''
  in \emph{Joint european conference on machine learning and knowledge
  discovery in databases}.\hskip 1em plus 0.5em minus 0.4em\relax Springer,
  2014, pp. 437--452.

\bibitem{MC91}
G.~A. Miller and W.~G. Charles, ``Contextual correlates of semantic
  similarity,'' \emph{Language and cognitive processes}, vol.~6, no.~1, pp.
  1--28, 1991.

\bibitem{DLT14}
W.~Ding, P.~Liang, A.~Tang, and H.~Van~Vliet, ``Knowledge-based approaches in
  software documentation: A systematic literature review,'' \emph{Information
  and Software Technology}, vol.~56, no.~6, pp. 545--567, 2014.

\bibitem{PXZ12}
R.~Pandita, X.~Xiao, H.~Zhong, T.~Xie, S.~Oney, and A.~Paradkar, ``Inferring
  method specifications from natural language api descriptions,'' in
  \emph{Proceedings of the 34th International Conference on Software
  Engineering}.\hskip 1em plus 0.5em minus 0.4em\relax IEEE Press, 2012, pp.
  815--825.

\bibitem{XYX16}
B.~Xu, D.~Ye, Z.~Xing, X.~Xia, G.~Chen, and S.~Li, ``Predicting semantically
  linkable knowledge in developer online forums via convolutional neural
  network,'' in \emph{Proceedings of the 31st IEEE/ACM International Conference
  on Automated Software Engineering}.\hskip 1em plus 0.5em minus 0.4em\relax
  ACM, 2016, pp. 51--62.

\bibitem{FAV18}
S.~Fakhoury, V.~Arnaoudova, C.~Noiseux, F.~Khomh, and G.~Antoniol, ``Keep it
  simple: Is deep learning good for linguistic smell detection?'' in \emph{2018
  IEEE 25th International Conference on Software Analysis, Evolution and
  Reengineering (SANER)}.\hskip 1em plus 0.5em minus 0.4em\relax IEEE, 2018,
  pp. 602--611.

\bibitem{MCC18}
M.~M{\"a}ntyl{\"a}, F.~Calefato, and M.~Claes, ``Natural language or not
  (nlon)-a package for software engineering text analysis pipeline,'' in
  \emph{2018 IEEE/ACM 15th International Conference on Mining Software
  Repositories (MSR)}.\hskip 1em plus 0.5em minus 0.4em\relax IEEE, 2018, pp.
  387--391.

\bibitem{TP10}
Y.~R. Tausczik and J.~W. Pennebaker, ``The psychological meaning of words: Liwc
  and computerized text analysis methods,'' \emph{Journal of language and
  social psychology}, vol.~29, no.~1, pp. 24--54, 2010.

\bibitem{KM18}
Z.~Kurtanovi{\'c} and W.~Maalej, ``On user rationale in software engineering,''
  \emph{Requirements Engineering}, vol.~23, no.~3, pp. 357--379, 2018.

\bibitem{RGQ12}
B.~Rogers, J.~Gung, Y.~Qiao, and J.~E. Burge, ``Exploring techniques for
  rationale extraction from existing documents,'' in \emph{2012 34th
  international conference on software engineering (ICSE)}.\hskip 1em plus
  0.5em minus 0.4em\relax IEEE, 2012, pp. 1313--1316.

\bibitem{HS09}
G.~E. Hinton and R.~R. Salakhutdinov, ``Replicated softmax: an undirected topic
  model,'' in \emph{Advances in neural information processing systems}, 2009,
  pp. 1607--1614.

\bibitem{ZS18}
Z.~Zhang and M.~Sabuncu, ``Generalized cross entropy loss for training deep
  neural networks with noisy labels,'' in \emph{Advances in Neural Information
  Processing Systems}, 2018, pp. 8792--8802.

\bibitem{MH17}
M.~C. Mukkamala and M.~Hein, ``Variants of rmsprop and adagrad with logarithmic
  regret bounds,'' in \emph{Proceedings of the 34th International Conference on
  Machine Learning-Volume 70}.\hskip 1em plus 0.5em minus 0.4em\relax JMLR.
  org, 2017, pp. 2545--2553.

\bibitem{DR10}
B.~Dagenais and M.~P. Robillard, ``Creating and evolving developer
  documentation: understanding the decisions of open source contributors,'' in
  \emph{Proceedings of the eighteenth ACM SIGSOFT international symposium on
  Foundations of software engineering}.\hskip 1em plus 0.5em minus 0.4em\relax
  ACM, 2010, pp. 127--136.

\end{thebibliography}
\end{document}